\documentclass{aa}  

\usepackage[normalem]{ulem}  
\usepackage{graphicx}  
\usepackage{txfonts}   
\usepackage{xcolor}    
\usepackage{orcidlink}
\usepackage{multirow}
\usepackage{hyperref}  
\usepackage{float}

\hypersetup{
    colorlinks=true,
    urlcolor=blue,
    linkcolor=blue,
    citecolor=blue
}
\def \MSUN{\rm M_{\odot}}  
\newcommand{\mytilde}{\raise.19ex\hbox{$\scriptstyle\sim$}} 
\begin{document}

\title{From Voids to Clusters: Mergers and Evolutionary Pathways of Star-Forming and Quenched Low-Mass Galaxies  }
\titlerunning{From Voids to Clusters}  

\author{
    Mohammad Reza Shojaei\textsuperscript{1}\,\orcidlink{0009-0004-7055-2203} \and
    Saeed Tavasoli\textsuperscript{1}\,\orcidlink{0000-0003-0126-8554} \and
    Parsa Ghafour\textsuperscript{1}\,\orcidlink{0009-0003-2960-1563}
}

\institute{
\textsuperscript{1}Department of Astronomy and High Energy Physics, Kharazmi University, No. 43 South Mofateh St, Tehran, Iran \\
\email{mrshojaei@ipm.ir} \\
}
\date{Received XX; accepted XX}

\abstract
{The evolution of low-mass galaxies is shaped by both internal processes and environmental factors, yet the role of environment and mergers in regulating their growth and star formation rates remains poorly understood, especially in the low-density regime.}
{This study aims to compare the evolutionary pathways and merger histories of star-forming and quenched galaxies in dense (cluster) and under-dense (void) environments, focusing on galaxies with stellar masses in the range \(10^{8.5} \leq M_\star/M_\odot \leq 10^{10.5}\). It presents the first statistical analysis within this stellar mass range, explicitly distinguishing between mini, minor, and major mergers across varying environments.}
{Using the high-resolution TNG300-1 simulation from the IllustrisTNG project, we classify galaxies as star-forming and quenched based on sSFR and UVJ criteria. We track their physical properties over the last \(\sim 10.5\) Gyr (\(z < 2\)), follow their complete merger histories, distinguishing between major, minor, and mini mergers, and assess the statistical impact of these mergers on star formation and gas content}
{At $z=0$, quenched galaxies in voids exhibit higher dark matter halo masses within $2\,R_{1/2}$ and retain more gas than their cluster counterparts, particularly at low stellar masses, while star-forming galaxies show similar dark matter and gas densities in both environments. Quenched systems experience a decline in SFR below $z\approx0.5$, whereas star-forming galaxies maintain relatively steady activity. Quenched galaxies assemble earlier than star-forming galaxies at all masses. Environmental effects are strongest for quenched galaxies at high mass, where cluster systems reach their final mass earlier than void systems. Quenched galaxies experience their last mergers systematically earlier than star-forming galaxies in both environments, consistent with their earlier mass assembly; however, the void-cluster environmental dependence of this timing differs markedly between the two populations, remaining small and mass-independent for star-forming galaxies, while showing a substantially larger, mass-dependent effect for quenched galaxies that reverses at the highest stellar masses. Merger activity shows a strong environmental dependence, quantified via the void-to-cluster merger-fraction ratio $R$: void star-forming galaxies experience more recent mini and minor mergers ($R\sim1.6$--$1.8$), while void quenched galaxies exhibit significantly higher late-time mini and minor merger ratios ($R\sim3.0$--$3.3$ and $R\sim2.7$--$2.9$). Mini mergers consistently enhance the star formation rate (SFR) and star formation efficiency (SFE), whereas minor and major mergers become effective mainly at $t_{\mathrm{lb}}\lesssim6\,\mathrm{Gyr}$. Major mergers drive the strongest increase in gas fraction in low-mass galaxies, while minor and major mergers contribute at comparable levels in high-mass systems. The characteristic stellar mass of merger companions (MeanStellarMass) also shows environmental trends: star-forming galaxies exhibit nearly identical distributions in voids and clusters, while quenched galaxies show a larger, though only marginally significant, shift toward higher merger masses in clusters.
Overall, merger-driven effects depend on both stellar mass and environment, with the strongest signatures occurring in low-mass galaxies in low-density environments.}
{} %

 \maketitle
%
\section{Introduction}
From large galaxy surveys, it is well established that the large-scale structure can be characterized as a 3D cosmic web with thin filaments connected by galaxy clusters and sheets surrounding under-dense regions \citep{bond1996filaments}. The current cosmological paradigm, $\Lambda$CDM, describes the origin of the cosmic web as the result of the formation and evolution of primordial density perturbations superposed on a homogeneous and isotropic background. In recent decades, substantial efforts have been focused on comprehending how galaxies evolve in different environments, ranging from dense regions such as groups and clusters to under-dense areas such as voids \citep{kauffmann2004environmental, peng2010mass}.

Two exciting aspects of how galaxy properties depend on their environments are star formation rates (SFRs) and morphology. Galaxies in denser regions, such as groups and clusters, exhibit attenuated star formation rates (SFRs), resulting in significantly higher fractions of red-sequence galaxies. Additionally, galaxies in these dense environments exhibit more elliptical morphologies than those in less dense environments, such as voids \citep{oemler1974systematic, davis1976galaxy,  dressler1980galaxy, postman1984morphology}. In particular, the environmental impact on SFR (or color) is more substantial than its influence on morphology, since changes in star formation activity occur more rapidly in response to environmental factors than structural transformations \citep{blanton2005relationship, ball2008galaxy, bamford2009galaxy}. 

Observations reveal that galaxies with stellar masses larger than $10^{10.5} \, M_{\odot}$ consistently exhibit low star formation rates (specific SFR $\sim 10^{-11} \, \text{yr}^{-1}$), regardless of their environment. This phenomenon, known as “mass quenching”, highlights the critical role of stellar mass in suppressing star formation \citep{peng2012mass, man2018star}. In this case, all internal processes of galaxies (secular evolution), including outflows from stellar winds, supernova explosions, and active galactic nucleus (AGN) feedback, contribute to the decrease in SFRs \citep{di2005energy, wake2012revealing}. On the other hand, in a high-density environment, the SFRs of low-mass galaxies whose stellar mass is smaller than $10^{10.5} \, M_{\odot}$ is also suppressed, defined as “environmental quenching”, attributes the lack of star formation in cluster galaxies to a variety of processes
that cause galaxies to lose their gas envelopes as they enter the cluster \citep{jaffe2016budhies, crossett2017near, medling2018sami, delgado2022minijpas}. In this case, various mechanisms halt the star formation of galaxies  \citep[for a recent review see e.g.][]{cortese2021dawes} : Ram pressure stripping results from interactions between
the hot cluster gas and the gas in an infalling galaxy, which removes
gas from the galaxy and thus stops star formation within about 1 Gyr \citep{gunn1972infall,  boselli2022ram}, especially in the center of clusters, tidal stripping \citep{merritt1984relaxation} which is a gravitationally driven process in which mass at the edges of an infalling galaxy is pulled off, strangulation or starvation \citep{larson1980evolution, balogh2000origin}, and harassment \citep{gallagher1972note}. These quenching mechanisms have been consistently proposed to describe the observed properties of passive galaxies \citep{contini2020roles, li2020characteristic}.

Cosmic voids constitute the most under-dense regions in the universe. Identifying these regions is non-trivial because it depends on how they are defined in shape, density thresholds, and the tracers used. The population of void galaxies tends to be dominated by low-mass, blue, gas-rich, star-forming galaxies with young stellar populations \citep{rojas2004photometric, hoyle2005luminosity, hoyle2012photometric, kreckel2014void, tavasoli2015galaxy, moorman2016star, florez2021void, jian2022star}. One possible reason for this is that galaxies in these surroundings may experience weaker gravitational forces due to the reduced density of matter, which could result in slower star formation histories (SFHs) compared to galaxies residing in denser environments, such as clusters, filaments, and walls \citep{dominguez2023galaxies}. Another possible consequence of residing in void environments is that these galaxies should be more solitary, have fewer nearby galaxies, and have less gas available for star formation. In addition, halos within the voids exhibit a slower evolution rate than those in denser environments \citep{rodriguez2022imprints}. Thus, these galaxies might exhibit unique morphologies and chemical compositions compared to denser areas. Void galaxies remain unaffected by transformation
processes (like ram-pressure stripping, starvation, and harassment)
that occur in group and cluster environments \citep{argudo2024morphologies, rodriguez2024evolutionary}. Consequently, by examining the characteristics of galaxies in cosmic voids, we can gain enhanced insights into the formation and evolution of galaxies, along with the large-scale structure of the universe.

Besides the considerable environmental influence on galaxy star formation rates, studies have underscored the significant impact of the interaction with other galaxies (i.e., major mergers)  in shaping star formation activity  \citep{di2005energy, croton2006many, hopkins2008cosmological, somerville2008semi}.
Observations support the picture that merger-induced instabilities drive gas flows into the center of merger remnants (\cite{blumenthal2018go}), dilute the gas-phase metallicity \citep{rupke2010galaxy, torrey2012metallicity, bustamante2018merger}, and enhance the star formation rate (SFR) on either nuclear or global scales \citep{ellison2008galaxy, robotham2014galaxy, sparre2022gas}. Numerical simulations of galaxy pairs provide an ideal framework to clarify the merger process and its effects on the galaxies involved, confirming that these interactions and mergers increase the star formation rate (SFR) \citep{di2007star,  cox2008effect, scudder2012galaxy, perret2014evolution, moreno2015mapping, moreno2021spatially} and regulate the gas content \citep{mihos1995gasdynamics, sinha2009numerical, sparre2022gas} in the nuclear region \citep{moreno2021spatially}. Research indicates that the typical increase in star formation rates (SFRs) of a merger is, at most, a factor of two, much lower than what would typically be considered a starburst \citep{ellison2013galaxy, knapen2015interacting, silva2018galaxy}. Mergers can potentially enhance the activity of an active galactic nucleus \citep[e.g.][]{sanders1996luminous, ellison2019definitive}. However, more recent studies indicate that this may not necessarily be true in every instance \citep[e.g.][]{darg2010galaxy, weigel2018fraction}.

Gas-rich (wet) mergers can sustain higher star formation rates because there is ample fuel for producing new stars \citep{lin2008redshift, perez2011chemical, athanassoula2016forming}. However, gas-poor (dry) mergers lack readily available gas, making it harder to form starbursts in these systems \citep{bell2006dry, naab2006properties, lin2008redshift}. In dense environments, dry galaxy mergers are more common than wet mergers because they are typically gas-poor than gas-rich galaxies \citep{lin2010wet}. The percentage of dry mergers also increases as the age of the Universe increases \citep{lin2008redshift}. Gas-poor galaxies dominate at high masses (stellar mass \(\gtrsim 10^{10.7} \MSUN\)), and as a result, mergers involving two high-mass galaxies usually result in dry interactions, which can suppress star formation \citep{robotham2014galaxy}.

Numerous prior studies have sought to identify the key factors that influence star formation variability in mergers, including the stellar mass ratio of the two galaxies, their initial orbital parameters, projected separation, total stellar mass, bulge to total mass ratio, and environment \citep{di2007star, cox2008effect, ellison2008galaxy, scudder2012galaxy, torrey2012metallicity,  perret2014evolution, domingue2016major, pan2018effect, jia2021relations}. The mass ratio of stellar mergers is a crucial parameter used to differentiate between major and minor mergers, traditionally set at 0.25 \citep{robotham2013galaxy, pan2018effect}. Designating the more massive member as the primary galaxy and the less massive member as the secondary, \citet{davies2015galaxy} found that primary galaxies enhance their star formation rate, regardless of whether they experience major or minor mergers. 

The study by \citet{sol2006effects} , utilizing the 2-degree Field Galaxy Redshift Survey (2dFGRS) and the Sloan Digital Sky Survey (SDSS), demonstrated that galaxy interactions are particularly effective at triggering significant star formation activity in low- and moderate-density environments. They found that the enhancement of star formation in major galaxy pairs is notably higher in low-density environments. Therefore, examining how galaxy mergers influence the star formation rates of galaxies in dense and under-dense environments is essential. 

Understanding the importance of mergers in the assembly history of galaxies requires studying galaxy merger rates as a function of cosmic time \citep{carlberg2000caltech, patton2002dynamically,  conselice2003evidence, lin2004deep2,lin2008redshift, lotz2008evolution, de2009vimos, bluck2009surprisingly} and understanding the level of triggered star formation during galaxy interactions \citep{lambas2003galaxy, nikolic2004star, woods2007minor, lin2007aegis, barton2007isolating}.

One of the main challenges in merger studies is the difficulty of identifying a large sample of merging galaxies. Visually detecting galaxies is time-consuming and often yields inconsistent results. Different observers may classify the same galaxy in various ways, and even the same observer might assign different labels on other days. Different mergers have been simulated, allowing us to study the SFRs of the merging galaxies throughout the entire merger sequence from the first passage to coalescence\citep[e.g.][]{springel2005modelling, hopkins2006unified, randall2008constraints, rupke2010galaxy}. These simulations have shown that SFR is enhanced when the merging galaxies are close to one another at first pass, second pass, and coalescence \citep{moreno2019interacting}. 

In recent years, it has become possible to make large-volume, cosmological and hydrodynamical simulations of galaxy formation and evolution with statistically significant and increasingly more realistic galaxy populations, using, for example, the EAGLE (Evolution and Assembly of Galaxies and their Environments) simulation suite \citep{crain2015eagle, schaye2015eagle}, the Illustris simulations \citep{vogelsberger2014introducing, vogelsberger2014properties, genel2014introducing}, Simba \citep{dave2019simba} or IllustrisTNG \citep{springel2018first, pillepich2018simulating}. These simulations, which employ a comprehensive galaxy formation model and state-of-the-art numerical code, are successful in reproducing a wide range of observations, including the cosmic star formation history, stellar population properties, stellar mass functions, scaling relations, clustering properties, galaxy sizes, and morphologies \citep{furlong2015evolution, rodriguez2015merger, sparre2017star, pillepich2018simulating, nelson2019illustristng}.

Despite numerous studies on galaxy interactions, particularly between massive galaxies, which emphasize their significant role in shaping galactic evolution \citep{patton2020interacting, brown2023interacting, byrne2023interacting}, studies of low-mass systems are relatively few and have predominantly focused on individual systems and have not thoroughly investigated the effects of mergers on low-mass (stellar masses $<10^{10.5} \, M_{\odot}$) evolution across environments ranging from high-density regions to under-dense voids.  Because low-mass galaxies are more susceptible to environmental influences than massive galaxies, it is crucial to investigate the effects of low-mass galaxy mergers on their evolution across diverse environments. 

IllustrisTNG provides a series of cosmological, gravity + magnetohydrodynamics simulations of galaxies in representative portions of synthetic universes carried out in a periodic box of 302.6 Mpc on a side \citep{springel2018first, marinacci2018first, nelson2018first,naiman2018first, pillepich2018simulating}. Because of the large volume covered by the simulation, the self-consistent treatment of baryons, and the physically motivated galaxy formation model used \citep{vogelsberger2013model}, the IllustrisTNG Simulation provides a unique opportunity to study the galaxy-galaxy merger rate with unprecedented precision and physical fidelity. 

In this study, we utilize the IllustrisTNG simulation to compare the evolutionary pathways and merger rate histories of low-mass star-forming and quenched galaxies in dense environments (groups and clusters) to those in under-dense voids. Although mergers in low-mass galaxies across different environments have received limited attention in previous research, this paper presents the first comprehensive statistical analysis of merger rates—distinguishing between mini, minor, and major mergers—in both dense and under-dense environments across various evolutionary stages of star-forming and quenched low-mass galaxies. Additionally, we examine the role of mergers in influencing the star formation histories of these low-mass galaxies. We also use the “Merger History” catalog \citep{rodriguez2017role} and \citep{eisert2023ergo}, which contains information and statistics on the merging history of all subhalos (i.e., galaxies) across time.

This paper is organized as follows. Section~2 provides an overview of the IllustrisTNG simulations and details the methodology used to identify galaxies in clusters and voids, classify them as quenched or star-forming, and apply thresholds based on specific star formation rates and the UVJ diagram. In Section~3, we present the results of our analysis, including the evolutionary behavior of galaxy properties, the mass assembly histories, and merger statistics. Finally, Section~4 summarizes the main findings and concludes the paper.

\begin{figure}[t] 
    \centering
    \includegraphics[width=0.5\textwidth]{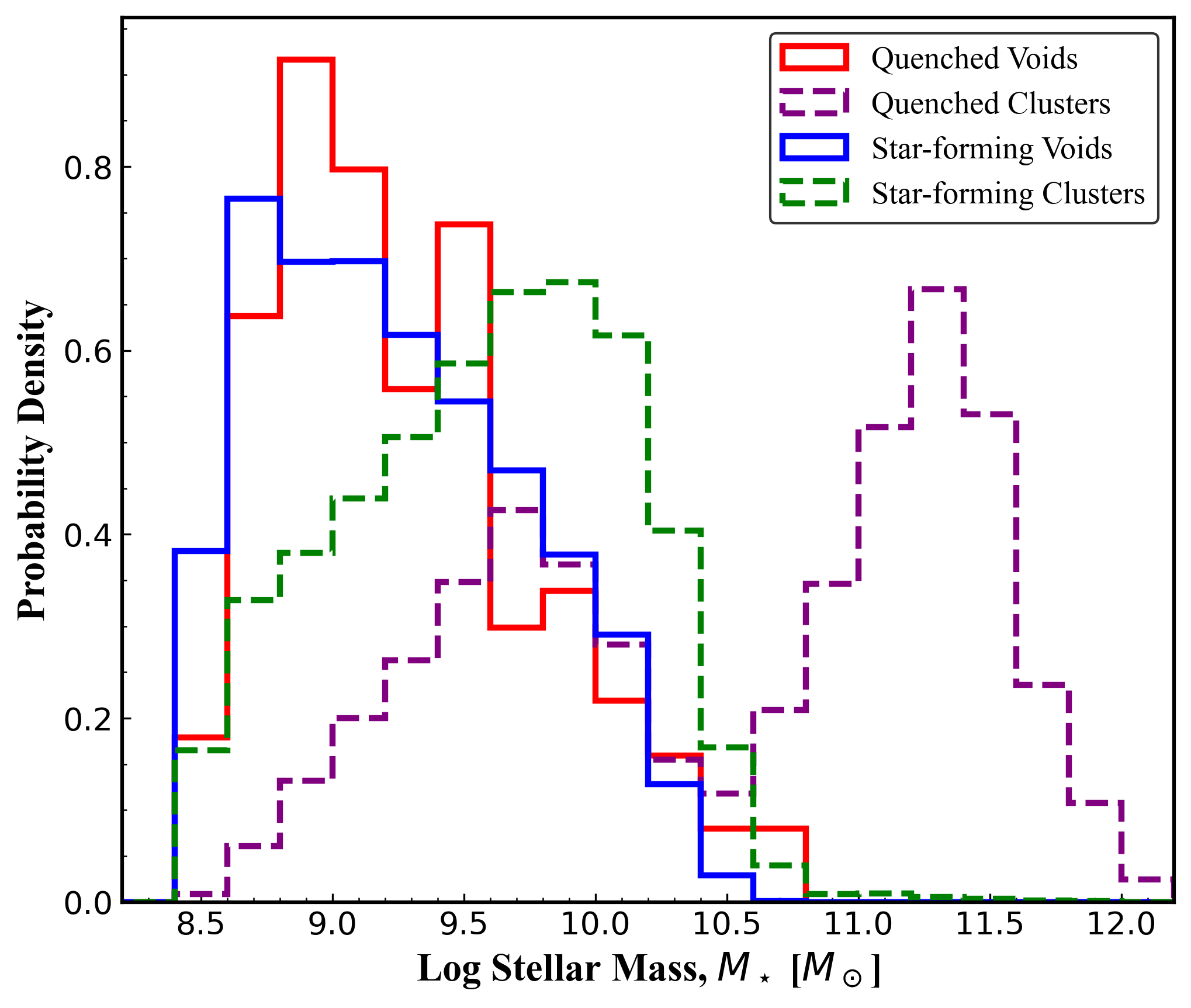} 
   \caption{Stellar-mass distributions of star-forming and quenched galaxies in void and cluster environments at $z=0$ after applying the sSFR and UVJ selection criteria.}

    \label{fig:mass_diagram} 
\end{figure}
\section{Methods}
\subsection{The IllustrisTNG simulations}
The IllustrisTNG simulations \citep{naiman2018first,  marinacci2018first, springel2018first, pillepich2018simulating, nelson2018first}, referred to as TNG, are a series of advanced cosmological magnetohydrodynamical simulations designed to investigate key physical processes vital for galaxy formation and evolution. These simulations build and improve upon the original Illustris project \citep{vogelsberger2013model,  torrey2015synthetic,  vogelsberger2014introducing, genel2014introducing, sijacki2015illustris, nelson2015illustris}, modifying and adding numerous features to improve the agreement between the simulation and observational results. They provide a comprehensive physical model of galaxy formation from the early Universe to the present \citep{weinberger2018supermassive, pillepich2018first}. The simulation suite contains three volumes: TNG50, TNG100, and TNG300, with sizes of \(51.7^3\), \(110.7^3\), and \(302.6^3\) comoving Mpc\(^3\), respectively. The TNG simulations adopt the cosmological parameters from Planck 2016, assuming \(\Omega_{\Lambda,0} = 0.6911\), \(\Omega_{m,0} = 0.3089\), \(\Omega_{b,0} = 0.0486\), \(\sigma_8 = 0.8159\), \(n_s = 0.9667\), and \(h = 0.6774\). 

This work focuses on results from the highest-resolution version of \texttt{TNG300\_1}, in which the dark matter (DM) and stellar particle/mean baryonic gas cell masses are \(5.9 \times 10^7\) and \(1.1 \times 10^7 \, \MSUN\), respectively. TNG300 has \(2 \times 2500^3\) initial resolution elements with a gravitational softening length of 1.48 kpc at \(z = 0\). This represents approximately an order-of-magnitude decrease in mass resolution and a factor of two increase in gravitational softening length compared to \texttt{TNG100\_1}.
The simulations are performed with the AREPO code \citep{springel2010moving}, which solves Poisson’s equation for gravity by employing a tree-particle-mesh algorithm. The code uses the finite volume method for magnetohydrodynamics in the simulation domain's unstructured, moving Voronoi tessellation. Poisson’s gravity equation is solved by employing a Tree-Particle Mesh \citep[TREE-PM;][]{xu1994new, bode2000tree, bagla2002treepm} that computes the contribution of short—and long-range forces using its tree and particle-mesh algorithms, respectively. Voronoi gas cells are particles at their center of mass, along with all other matter components. In the IllustrisTNG simulation, halos and their associated subhalos are identified using the Friends-of-Friends \citep[FoF;][]{davis1985evolution} and SUBFIND \citep{springel2001populating} algorithms, respectively. Within each FoF halo, subhalos identified by SUBFIND consist of all resolution elements—gas, stars, dark matter, and black holes—which are gravitationally bound to the subhalo. The merger trees we used here were constructed with SUBLINK \citep{rodriguez2015merger}. A subhalo's descendant is the one that shares the highest weighted sum of individual particles—gas, stars, and dark matter—with its progenitor. These particles are ranked by gravitational binding energy and weighted by $(\text{rank})^{-1}$. A merger occurs when multiple subhalos have a shared descendant, with the main progenitor identified as the subhalo with the most massive history \citep{de2006formation}.

\subsection{Merger History in the TNG}
In this research, we use catalogues of merger histories from \citet{rodriguez2017role} and \citet{eisert2023ergo}, which include data and statistics on the merger history of all subhalos (i.e., galaxies) across cosmic time. Here, mergers are split into three categories: major (stellar mass ratio > 1/4), minor (stellar mass ratio between 1/10 and 1/4), and mini (stellar mass ratio < 1/10) mergers. The mass ratio is always determined by the stellar masses of the two merging galaxies when the secondary galaxy reaches its maximum stellar mass. To avoid spurious flyby and re-merger events, mergers are only included when both galaxies can be traced back to a time when each belonged to a different FoF group. We also remove subhalos identified as non-cosmological structures in the TNG300 group catalog (\texttt{SubhaloFlag} = 0). Such systems generally correspond to poorly resolved or numerically spurious objects and are not considered physically meaningful galaxies.

\begin{figure}[t] 
    \centering
    \includegraphics[width=0.5\textwidth]{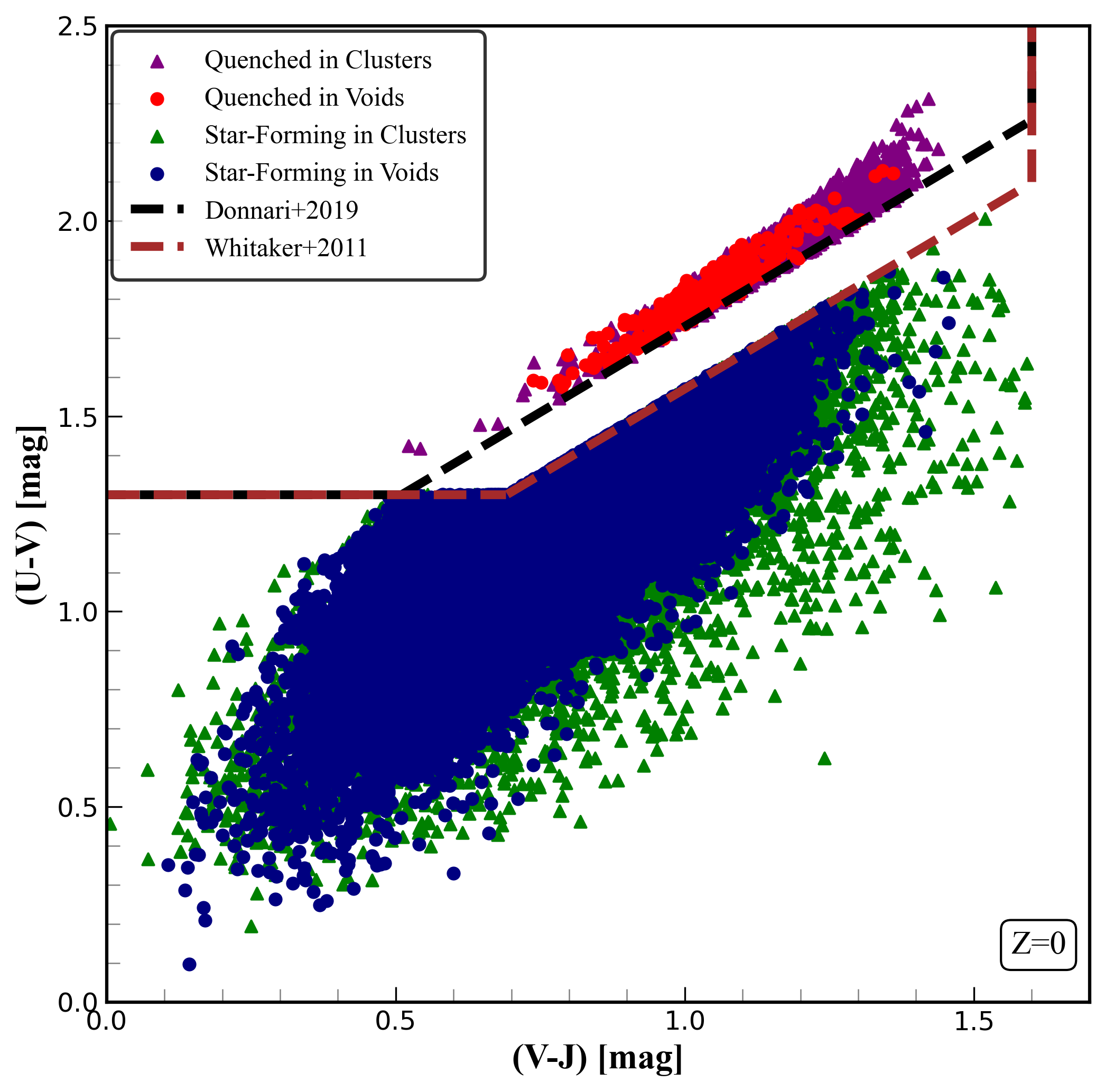} 
   \caption{The UVJ diagram at $z=0$ shows the final galaxy samples used in this work: star-forming (blue circles) and quenched (red circles) galaxies in voids, and star-forming (green triangles) and quenched (purple triangles) galaxies in clusters. The samples are defined by applying the sSFR and UVJ selection criteria, together with the condition $R_{\mathrm{cluster\text{-}centric}} < 3R_{200}$ for group and cluster galaxies, and restricting the stellar-mass range to $10^{8.5} \leq M_{\star}/M_{\odot} \leq 10^{10.5}$. The brown dashed line indicates the quiescent selection threshold from \citet{whitaker2011newfirm}, while the black dashed line corresponds to the criterion from \citet{donnari2019star}, adopted for the TNG (IllustrisTNG) simulations.}

    \label{fig:UVJ_diagram} 
\end{figure}

\subsection{Galaxy sample}
\subsubsection{Finding galaxies}
This paper uses the publicly available data from the most oversized simulation box \citep{nelson2019illustristng} with the best resolution at this level, TNG300-1 (hereafter TNG300), in the $z = 0$ redshift. The properties of halos and subhalos utilized in this study are derived from the application of the Friends-of-Friends (FoF) and SUBFIND algorithm \citep{springel2001populating}, which are employed to detect substructures and, consequently, galaxies within the simulated volumes. In this study, we obtain the stellar masses by summing the masses of all stellar particles located within twice the stellar half-mass radius ($M_\star$). For consistency, all other galaxy properties considered in this work (e.g., gas mass, dark matter mass, gas number density) are computed using particles of the corresponding type within the same aperture (twice the stellar half-mass radius), unless stated otherwise. Considering numerical resolution effects in TNG300, we apply a stellar mass cut-off of \(10^{8.5} M_\odot\) for the low-mass end of our galaxy sample \citep{pillepich2018first}. In the first step, we select all subhalos that satisfy the criterion \(M_\star \geq 10^{8.5} \, M_\odot\) and identify $\sim 400,000$  galaxies in TNG300 at \(z = 0\).

We note that although the low-mass threshold is chosen based on the reliability of dark-matter subhalo tracking, the merger classification (major, minor, mini) is defined using stellar mass ratios \citep{rodriguez2017role}, which are more sensitive to resolution at the lowest masses considered here ($M_\star \sim 10^{8.5}\,M_\odot$, corresponding to $\sim 30$ stellar particles). We therefore emphasize that our results should be interpreted as most robust within stellar mass ranges resolved with a larger number of stellar particles, and caution that the exact merger statistics in the lowest mass bin may carry additional uncertainty from stellar-mass resolution.
 
\subsubsection{Group \& Cluster Galaxies}
The large volume of TNG300 enables the study of a significant number of simulated groups and clusters, allowing for comprehensive comparisons of galaxy properties in dense and under-dense environments \citep{pillepich2018first}. For galaxies residing in dense environments, we focus on structures with masses \(M_{200} \geq 10^{13} \, M_\odot\), containing at least ten galaxies per halo at \(z = 0\).\footnote{The radial physical scale \(R_{200}\) of FoF halos is defined as the radius of a sphere centered on the halo that encloses a mass \(M_{200}\) with a density equal to 200 times the critical density of the Universe.} These thresholds result in the identification of 2,920 groups with masses \(10^{13} \leq M_{200} < 10^{14} \, M_\odot\), 184 clusters with masses \(10^{14} \leq M_{200} < 10^{14.5} \, M_\odot\), and 26 clusters with masses \(10^{14.5} \leq M_{200} \leq 10^{15} \, M_\odot\). Approximately 101,500 galaxies with stellar masses \(M_\star \geq 10^{8.5} \, M_\odot\) are identified in these dense environments. To study the effects of dense environments on galaxy properties, we focus exclusively on galaxies residing within \(3R_{200}\) of each group and cluster, i.e., those with \(R_\text{ cluster-centric} < 3R_{200}\). Additionally, we exclude galaxies without star formation (\(\text{SFR} = 0\)). Applying these criteria, the final sample of galaxies in groups and clusters totals $\sim 28,600$.

\subsubsection{Void Identification}
Cosmic voids constitute the most under-dense regions in the universe. Identifying these regions is challenging because their definitions rely on factors such as shape, density thresholds, and the tracers used. The literature offers various ways to define them \citep[e.g.][]{kirshner1981million, kauffmann1991voids, sahni1994evolution, benson2003galaxy, padilla2005spatial, platen2007cosmic, neyrinck2008zobov, lavaux2010precision, sutter2015vide, ghafour2025vega}. Overall, they all exhibit similar characteristics for galaxies located within these voids. The populations of void galaxies are generally composed of low-mass, blue, star-forming galaxies with young stellar populations \citep{rojas2004photometric, hoyle2005luminosity, hoyle2012photometric, tavasoli2015galaxy, moorman2016star, florez2021void, jian2022star}.

For identified void galaxies from the initial data, we employed the void finder algorithm introduced by \citet{aikio1998simple}  (subsequently referred to as the AM algorithm), which has been enhanced to a three-dimensional version \citep{tavasoli2013challenge}. The AM void finder algorithm identifies under-dense regions in galaxy surveys or cosmological simulations. The method begins by discretizing the spatial distribution of galaxies into a 3D grid, where each cell’s density is estimated using a kernel-based approach or by counting nearby galaxies. Regions with densities below a specified threshold are flagged as potential void candidates. Adjacent low-density cells are grouped using a friends-of-friends algorithm, forming connected under-dense structures. To refine results, overlapping or proximate voids are merged iteratively, ensuring coherent void boundaries. The algorithm applies a minimum size criterion to exclude small, insignificant voids, emphasizing cosmologically relevant structures. Void identification is performed at $z=0$ using a volume-limited tracer sample of galaxies with stellar mass $M_\star > 10^{8}\,M_\odot$, which serve as tracers of the density field. For each identified void, we compute the luminosity density contrast defined as $\delta_v = (\rho_v - \rho_m)/\rho_m$, where $\rho_v$ is the total luminosity of galaxies inside the void divided by its volume, and $\rho_m$ is the mean luminosity density of the tracer sample. The effective radius $R_v$ of each void is defined as the radius of a sphere whose volume equals that of the void. To avoid spurious detections, we retain only voids with effective radii $R_v > 7\,\mathrm{Mpc}$.
The method efficiently handles large datasets by prioritizing computational simplicity, making it suitable for cosmological analyses. Despite its straightforward design, the Aikio-Mahonen algorithm effectively captures the hierarchical nature of void networks, providing insights into the universe's large-scale structure through tunable parameters like density thresholds and void size limits. Our final count of void galaxies with \(\text{SFR} \neq 0\) and stellar masses \(M_\star \geq 10^{8.5} \, M_\odot\) is $\sim 27,867$.

The distinction between quenched and star-forming galaxies at $z = 0$ is a significant area of research. It is often ambiguous and somewhat arbitrary from theoretical and observational perspectives. Therefore, in this study, we aim to make a substantial contribution by simultaneously exploring two different classification criteria, the UVJ Diagram and the Specific Star Formation Rate threshold, to differentiate these two populations of galaxies more effectively.

\subsubsection{Galaxy Classification Using the UVJ Diagram}
A significant method for identifying quiescent galaxies commonly employed in observations is the color-color diagram, especially the rest-frame U-V versus V-J plane. \citep[Hereafter, UVJ diagram;][]{wuyts2011star, williams2009detection, williams2010evolving, whitaker2010age, patel2013hst, quadri2011tracing}. For plotting the UVJ diagram, we use the “SDSS ugriz and UVJ Photometry/Colors with Dust” catalog in TNG300 at redshift \(z=0\). This catalog contains synthetic stellar photometry (i.e., colors), including the effects of dust obscuration, corresponding to the fiducial dust model of \citet{nelson2018first} (i.e., Model C). Specifically, we utilize the rest-frame UVJ catalog as analyzed by \citet{donnari2019star}. 

 We adopt two different thresholds to separate star-forming and quenched galaxies based on UVJ cuts; first, apply the criteria derived from \citet{whitaker2011newfirm}:
\[
(U - V ) > 0.88 \times (V - J) + 0.69, \quad \text{applied for} \; z \leq 0.5.
\]

Secondly, we utilize the best estimate from \citet{donnari2019star} to differentiate TNG galaxies in the UVJ diagram:
\[
(U - V ) > 0.88 \times (V - J) + 0.85, \quad \text{applied for} \; 0 \leq z \leq 1.
\]
In addition, for the limits on \( U - V \) and \( V - J \), we arbitrarily use the ones defined in  \citet{whitaker2011newfirm}: 
\[
(U - V ) > 1.3 \quad \& \quad (V - J) < 1.6, \quad \text{valid for} \; 0 < z < 1.5.
\]

\subsubsection{Specific Star Formation Rate Threshold}
In the TNG model, star formation follows the approach described by \citet{springel2003cosmological}. Gas is converted into star particles in a stochastic manner when its density exceeds \(n_H = 0.1\, \mathrm{cm^{-3}}\). This conversion occurs over a timescale empirically determined to align with the Kennicutt-Schmidt relation \citep{kennicutt1989star}. For each galaxy's star-formation rate (SFR), we use the sum of the instantaneous SFRs of all the gas cells assigned twice the stellar half-mass radius. In the literature, we commonly define "quenched" galaxies as those whose logarithmic Specific Star Formation Rate (sSFR) falls below a certain fixed threshold at any redshift, namely \(\mathrm{sSFR} \leq 10^{-11}\, \mathrm{yr^{-1}}\). Star-forming galaxies are thus those with sSFR larger than this threshold. Such threshold-based separation is often used in observations \citep{mcgee2011dawn, wetzel2013galaxy, lin2014environment, jian2018first}.

Applying the sSFR threshold to the void sample yields 27,351 star-forming and 516 quenched galaxies. Incorporating the UVJ criterion refines these numbers to 16,420 and 250 galaxies, respectively. For group and cluster galaxies, the sSFR threshold identifies 24,942 star-forming and 3,696 quenched systems, which are reduced to 13,812 and 2,874 galaxies after applying the UVJ selection.

We note that our adopted log(sSFR) = -11 threshold, together with the UVJ criterion, follows the definition introduced by \citet{donnari2019star} for TNG galaxies, rather than being imported solely from the observational literature. As shown by \citet{donnari2019star}, while TNG reproduces the observed bimodality in galaxy colors, the simulated SFR (sSFR) distributions are unimodal, peaked at the star-forming main sequence, unlike the bimodal sSFR distributions seen in observations. We retain the combined sSFR+UVJ criterion for consistency with this reference and to ensure a clean separation between quenched and star-forming populations, at the cost of a smaller quenched sample size.

Since galaxy properties strongly correlate with stellar mass ($M_{\star}$), environmental trends must be evaluated at fixed $M_{\star}$. Figure \ref{fig:mass_diagram}  shows the stellar-mass distributions of the final void and cluster samples after applying the sSFR and UVJ criteria. As expected, void galaxies are predominantly low-mass systems, while quenched cluster galaxies extend to higher stellar masses. To ensure meaningful comparisons and sufficient statistics in both environments, we restrict our analysis to the mass range $10^{8.5} \leq M_{\star}/M_{\odot} \leq 10^{10.5}$. All environmental comparisons are performed within fixed stellar-mass bins, thereby controlling for stellar-mass dependence without applying additional global mass-matching procedures.

\begin{table}[h!]
\centering
\caption{The final number of Star-Forming and Quenched Galaxies with a stellar mass range of \(10^{8.5} \leq M_\star \leq 10^{10.5} \, M_\odot\) for both void and cluster samples.}
\begin{tabular}{lc}
\hline
\textbf{Category}            & \textbf{Number of Galaxies} \\ \hline
Star-Forming in Voids        & 16,298\\
Quenched in Voids            & 246\\
Star-Forming in Clusters     & 13,449\\
Quenched in Clusters         & 1,324\\ \hline
\end{tabular}
\label{tab:final_samples}
\end{table}

 Figure~\ref{fig:UVJ_diagram} shows our final samples in void and cluster environments within the UVJ diagram at redshift \(z=0\). The diagram displays the final samples of star-forming (blue dots) and quenched (red dots) void galaxies, as well as star-forming (green triangles) and quenched (purple triangles) galaxies in groups and clusters. As detailed above, these samples were obtained by applying all thresholds on void and cluster galaxies with stellar masses in the range \(10^{8.5} M_\odot \leq M_\star \leq 10^{10.5} M_\odot\). The brown dashed line represents the classification threshold from \citet{whitaker2011newfirm}, while the black dashed line follows the classification threshold from \citet{donnari2019star}, used for TNG (IllustrisTNG). Table~\ref{tab:final_samples} also presents the final number of star-forming and quenched galaxies in voids and clusters, as depicted in Figure~\ref{fig:UVJ_diagram}, after applying all thresholds to the void and cluster samples at $z = 0$ in TNG300 simulation. 

\section{Results}
\subsection{Star-forming and quenched galaxies as a function of Stellar Mass}

\begin{figure*}[t] 
    \centering
    \includegraphics[width=\textwidth, keepaspectratio]{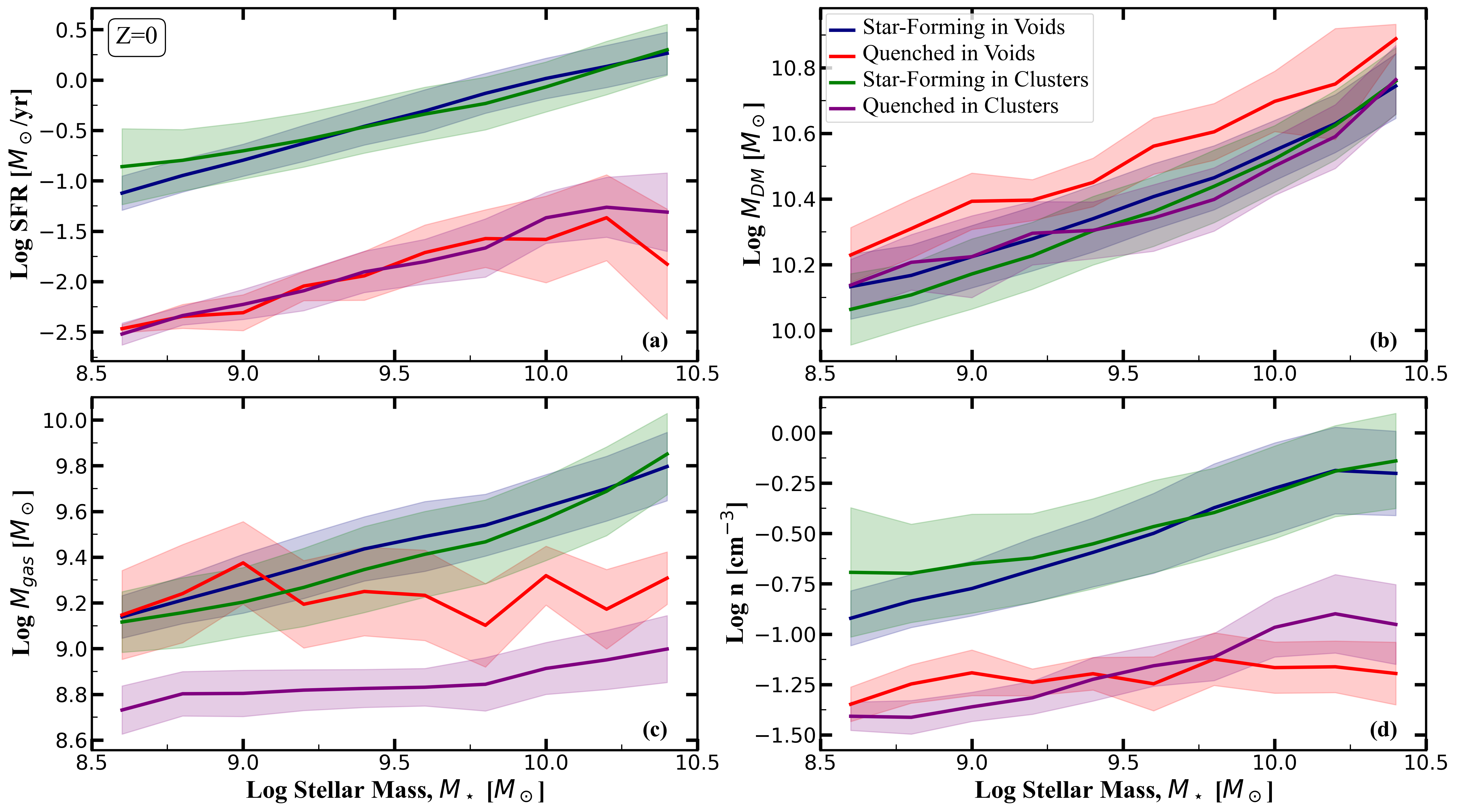 }
    \caption{Comparative analysis of six key parameters: star formation rates(SFRs), the mass of dark matter (\(M_\mathrm{DM}\)), the mass of gas (\(M_\mathrm{gas}\)) and the average number density of gas (\(n\)), all within twice the half-mass stellar radius for star-forming (blue lines) and quenched (red lines) void galaxies and star-forming (green lines) and quenched (purple lines) for cluster galaxies as a function of stellar mass in TNG300 at \(z=0\). The solid lines show the median values in each stellar-mass bin, and the shaded regions correspond to the median absolute deviation (MAD), which provides a robust measure of the intrinsic scatter.}
    \label{fig:stellar_mass_2comparison}
\end{figure*}

Before examining the evolutionary pathways and merger rate histories of low-mass star-forming and quenched galaxies in void and cluster environments, we compare the essential characteristics of these galaxies at \(z=0\). In Figure \ref{fig:stellar_mass_2comparison}, we present a comparative analysis of the properties of star-forming and quenched low-mass galaxies with stellar mass ($M_\star \)) ranging from \(10^{8.5} \, M_\odot\) to \(10^{10.5} \, M_\odot\) in dense and under-dense environments. We examine the logarithms of five parameters, all measured within twice the stellar half-mass radius of each galaxy: star formation rates (SFRs), dark matter mass ($M_\mathrm{DM}$), gas mass ($M_\mathrm{gas}$), specific star formation rate ($\mathrm{sSFR} \equiv \mathrm{SFR}/M_\mathrm{stars}$), and gas number density ($n$), as a function of stellar mass. The blue and red solid lines represent the median relation of each parameter versus stellar mass for star-forming and quenched void galaxies, respectively. In contrast, the green and purple solid lines denote the same for star-forming and quenched galaxies in clusters.
To quantify the dispersion, we also utilize shaded error bars to represent the median absolute deviation (MAD) around the median, defined as $\mathrm{MAD}=\mathrm{median}(|x_i-\mathrm{median}(x)|)$. We use the MAD as a robust estimate of the scatter within each bin.

The overall view of Figure \ref{fig:stellar_mass_2comparison}, indicates that star-forming galaxies, regardless of residing in voids or clusters, approximately follow consistent trends across all evaluated parameters. There are slight variations in dark matter mass between voids and clusters within specific stellar mass ranges. In contrast, quenched galaxies in voids and clusters show significant differences in their properties, highlighting that the quenching mechanisms in these two environments are fundamentally distinct. These differences are particularly evident in the gas mass across all stellar mass bins, highlighting the contrasting processes that drive quenching in voids and clusters. Given that our sample comprises low-mass galaxies, dense environments exert a more pronounced influence on quenched cluster galaxies than their void counterparts, emphasizing the impact of “environmental quenching” \citep{peng2010mass, cybulski2014voids}.

Figure \ref{fig:stellar_mass_2comparison} (a) effectively depicts the correlation between star formation rates (SFRs) and stellar mass ($M_\star \)), commonly referred to as the “main sequence” of star-forming galaxies, at \(z=0\) \citep{brinchmann2004physical, schreiber2015herschel}. It highlights that low-mass star-forming galaxies, regardless of their environments (voids or clusters), follow a consistent trend along the main sequence, emphasizing the universality of this relationship. Furthermore, quenched galaxies in two environments display a comparable SFR -$M_\star \) correlation with markedly lower SFRs, as determined by our thresholds to distinguish the sample's two low-mass populations (star-forming and quenched). 

Figure \ref{fig:stellar_mass_2comparison} (b) presents a detailed analysis of the correlation between the stellar mass of the sample galaxies and the dark matter mass of their host halos within twice the stellar half-mass radius, as characterized by the stellar mass–halo mass (SM–HM) relation for low-mass galaxies ($M_\mathrm{DM} < 10^{12}\, M_\odot$) \citep{behroozi2010comprehensive, moster2013galactic}. Notably, our findings confirm that the (SM–HM) relation holds across different environments for star-forming and quenched galaxies, regardless of environment. Interestingly, the figure reveals that quenched void galaxies reside in more massive dark matter halos across all stellar mass bins at redshift \(z=0\), compared to all of our samples of star-forming and quenched galaxies in voids and clusters.

Figure \ref{fig:stellar_mass_2comparison} (c) illustrates that star-forming galaxies, regardless of whether they reside in voids or clusters, exhibit a consistent and positive correlation between gas mass ($M_\mathrm{gas}$) and stellar mass ($M_*$). This correlation emphasizes that as stellar mass increases, these galaxies retain more gas to fuel ongoing star formation. Also, the discrepancy between quenched galaxies and star-forming galaxies increases in higher stellar mass bins. Additionally, in voids, quenched galaxies exhibit approximately no change in gas mass as stellar mass increases and show little to no positive correlation between gas mass and stellar mass. Interestingly, the figure shows that in the same stellar mass bins, the total gas mass is higher in quenched galaxies in voids than in clusters, particularly at low stellar masses, which can refer to gas stripping of low-mass galaxies in clusters. For example, ram pressure stripping, which strips gas from low-mass galaxies ($M_* < \ 10^{10.5} \, M_\odot$) more easily due to their shallow gravitational wells \citep{steinhauser2016simulations, cora2019semi, boselli2022ram}. Another possibility is that during the infalling process, their gas will be stripped off, which directly leads to their quenching \citep{bekki2002passive}. In this case, the effect should be stronger for these low-mass galaxies, as their gas components are more easily stripped.  

 The gas cell number density is calculated as the average of all gas cells located within twice the half-mass radius of each galaxy, as shown in Figure \ref{fig:stellar_mass_2comparison} (d). The figure reveals that star-forming galaxies in both voids and clusters exhibit higher gas cell number densities ($n$) than their quenched counterparts across all stellar mass bins, with a positive correlation between gas number density and stellar mass. This trend is also observed for quenched galaxies in clusters; however, the correlation is weaker for quenched galaxies in voids. Furthermore, within the stellar mass range $10^{8.5}-10^{9.5} \, M_\odot$, star-forming galaxies in clusters have slightly higher gas cell number densities than their counterparts in voids. Quenched void galaxies exhibit slightly higher gas cell number densities in this range than quenched cluster galaxies. However, this trend reverses for quenched galaxies at higher stellar masses. The multi-phase interstellar medium (ISM) star formation and pressurization are treated following the \citep{springel2003cosmological}  model in TNG simulations. Specifically, gas above a star formation density threshold of \(n > 0.1 \, \mathrm{cm}^{-3}\) forms stars stochastically following the empirically defined Kennicutt-Schmidt relation and assuming a Chabrier \citep{chabrier2003galactic} initial mass function. 
 
 Therefore, the presence of non-zero total gas mass in quenched galaxies does not necessarily imply ongoing star formation. Instead, our results indicate that quenched galaxies retain gas that is systematically less dense than in star-forming systems. This suggests that quenching in these galaxies is associated with the suppression of the dense star-forming gas phase rather than complete gas removal. In other words, although gas remains present, a smaller fraction of it exceeds the star-formation density threshold, preventing efficient star formation.
 
\subsection{Evolutionary Analysis of Galaxy Properties}
\begin{figure*}[ t]
    \centering
    
    \includegraphics[width=0.9\textwidth, keepaspectratio]{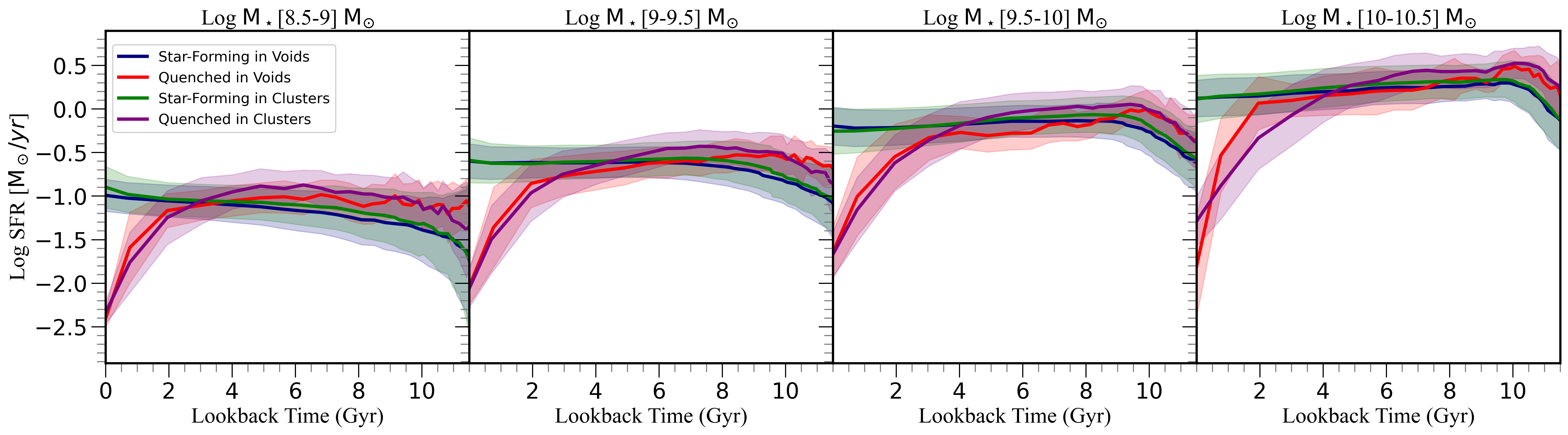}
    
    \vspace{0.3em} 
    
    \includegraphics[width=0.9\textwidth, keepaspectratio]{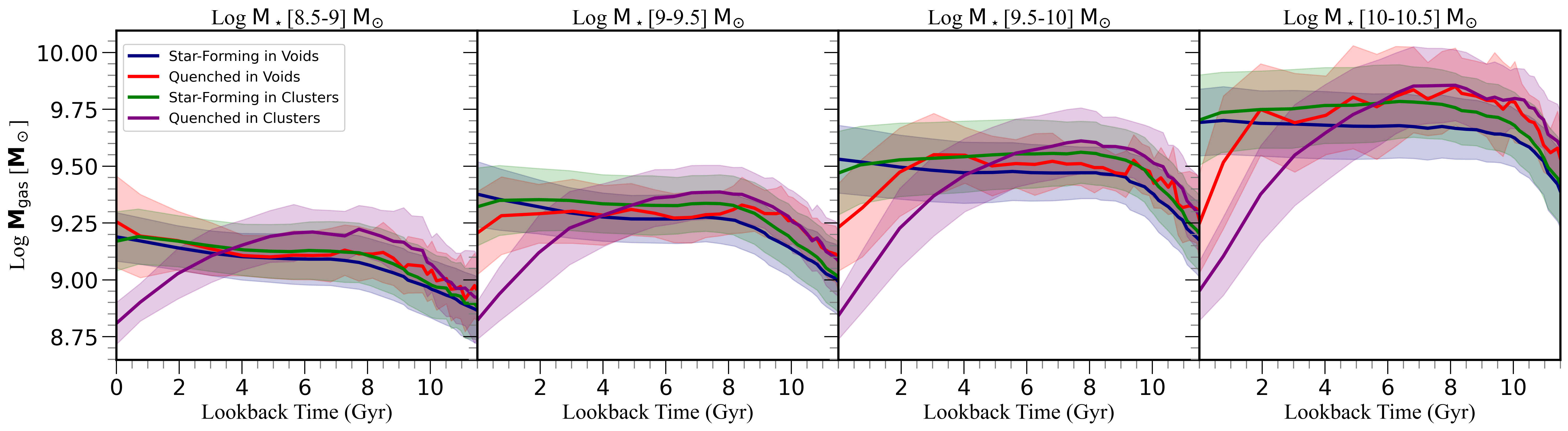}
    
    \vspace{0.3em} 
    
    \includegraphics[width=0.9\textwidth, keepaspectratio]{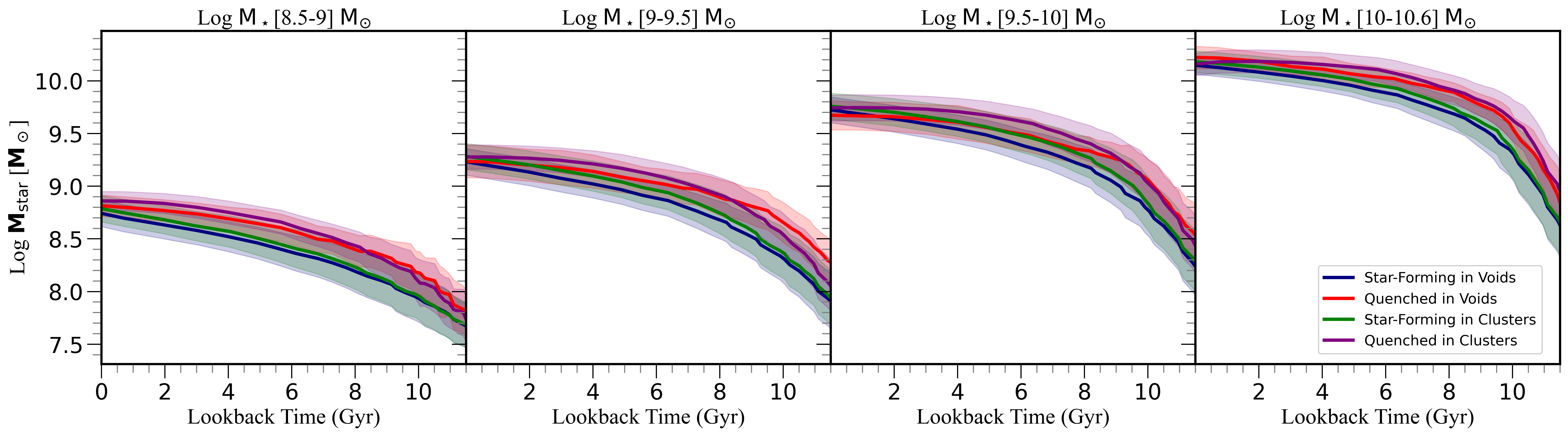}
    
    \vspace{0.3em} 
    
    \includegraphics[width=0.9\textwidth, keepaspectratio]{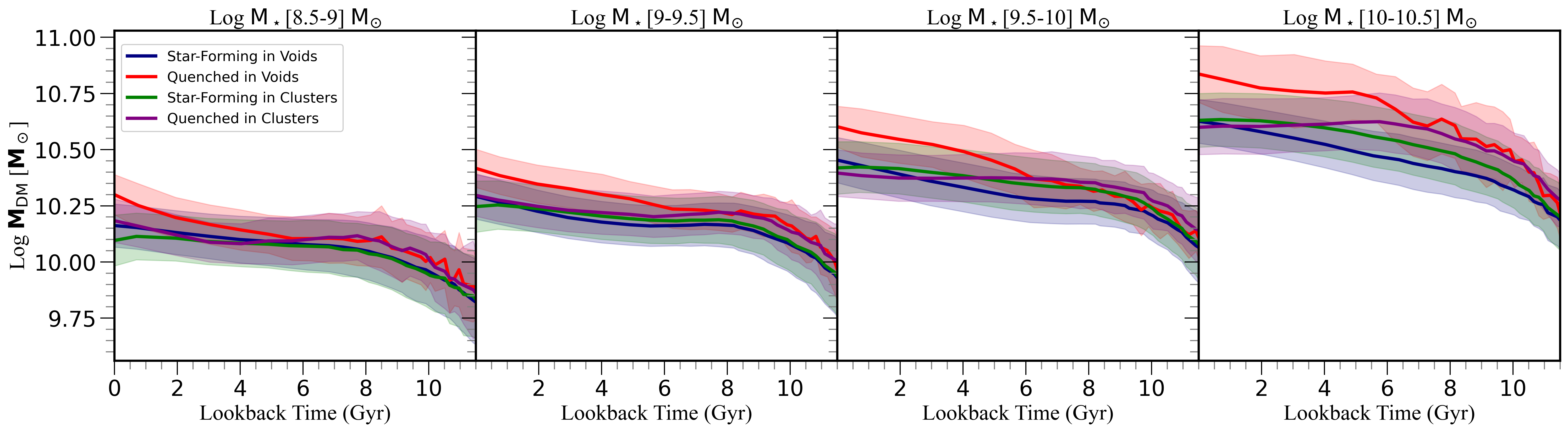}

        \vspace{0.3em} 
    
    \includegraphics[width=0.9\textwidth, keepaspectratio]{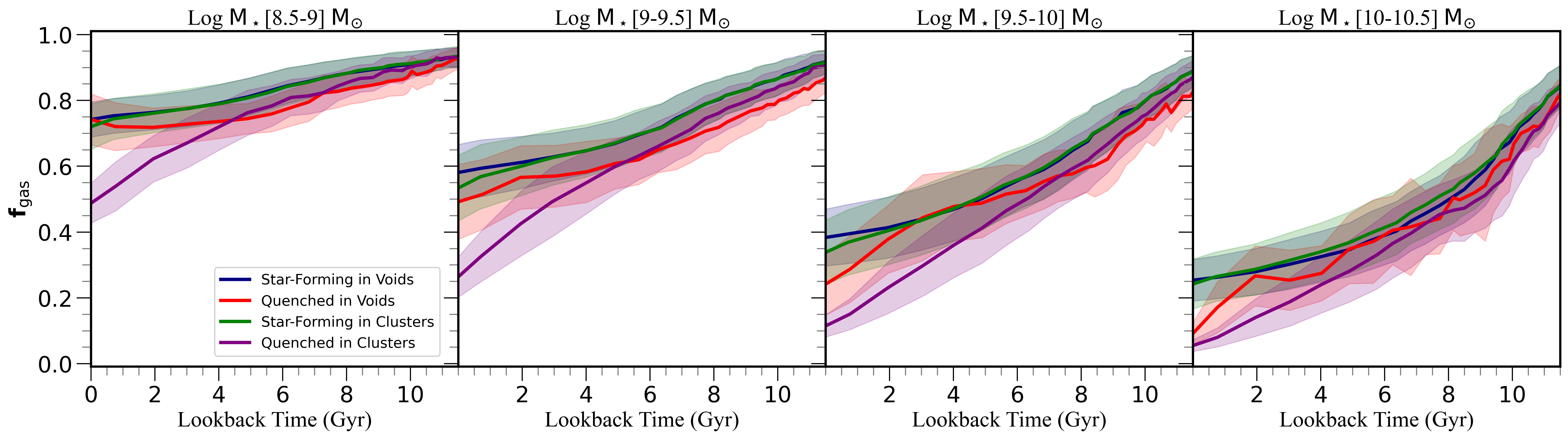}
    \caption{Evolution of star formation rates (SFRs), gas mass ($M_\text{gas}$), stellar mass ($M_\text{star}$), dark matter mass ($M_\text{DM}$)  and  gas fraction ($f_\text{gas}$) within twice the stellar half-mass radius, as a function of  \(lookback time \leq 10.5  Gyr\) (\(z \leq 2\) ) for star-forming (blue and green) and quenched (red and purple) galaxies in voids and clusters, respectively, in the IllustrisTNG 300 simulation.  Lines in each panel represent the median of each parameter within redshift bins, while the shaded regions indicate the absolute error on the median, reflecting measurement uncertainties.}
    \label{fig:01stellar_mass_comparison03}
\end{figure*}

The quenching of galaxies results from a complex interplay between stellar mass and environmental factors, with their influence varying based on galaxy mass and redshift. At high redshifts (\(z > 1\)), mass quenching predominantly affects massive galaxies \citep{peng2010mass}, while at lower redshifts (\(z \leq 1\)), environmental quenching becomes particularly significant
 \citep{balogh2016evidence, darvish2016effects}, especially for low-mass galaxies in dense regions where the environment plays a critical role \citep{peng2012mass}. Consequently, understanding the evolutionary trajectories of void and cluster galaxies is key to deciphering the processes governing galaxy formation and transformation across cosmic time in dense and under-dense environments. In Figure \ref{fig:01stellar_mass_comparison03} , we illustrate the evolution of key galaxy properties, such as star formation rates (SFRs),gas mass ($M_\text{gas}$), stellar mass ($M_\text{star}$), dark matter mass ($M_\text{DM}$),  , the gas fraction (\(f_{\rm gas}\equiv M_{\text{gas}}/(M_{\text{gas}} + M_{\text{star}}) \)), all within twice the stellar half-mass radius, provides insights into the underlying physical 
 
 \clearpage
\clearpage

mechanisms that drive galaxy growth and quenching in different regions. This section examines the evolutionary trends of star-forming and quenched galaxies in void and cluster environments, spanning a range of lookback times \(\tau \leq 10.5 \, \text{Gyr}\) (\(z \leq 2\)).
Each plot consists of four lines: blue and green lines represent the median values of star-forming galaxies in voids and clusters, respectively, while red and purple lines correspond to the median values for quenched galaxies in voids and clusters. Surrounding each curve, the shaded regions illustrate the median absolute deviation (MAD) within each redshift bin, offering a robust measure of variability across the data. 

Our overall results of the evolution of main parameters across lookback time show that low-mass star-forming galaxies behave independently of their environments and evolve similarly in dense and under-dense regions. Our results agree with the general understanding that the environment does not significantly influence the SFRs of star-forming galaxies \citep{lagana2018star, contini2019roles}. Our results reveal that, for star-forming galaxies, the median SFR are similar across different environments, regardless of redshift and stellar mass, in agreement with \citep{darvish2016effects}.  Also, the Figure indicates that the evolution of SFR in low-mass quenched galaxies, regardless of their environment, behaves approximately the same, and both populations decrease significantly over the last 5 Gyr ($z \sim 0.5$) despite exhibiting
higher SFR at higher redshifts. However, the reasons behind the quenching of these galaxies may differ. Additionally, other parameters of these quenched galaxies exhibit distinct behaviors, suggesting that different mechanisms might be at play in their evolution.  

Despite the relatively minor influence of the environment on star-forming galaxies, as shown in Figure \ref{fig:01stellar_mass_comparison03}, Quenched void and cluster galaxies exhibit substantial differences, particularly in their gas mass within twice the stellar half-mass radius. Quenched cluster galaxies experience significant gas mass depletion over the past 4 Gyr for the lower stellar mass bins and 6 Gyr for the higher stellar mass bins. These galaxies experience earlier and more efficient gas loss than their counterparts in voids, leading to lower final gas masses. This depletion likely results from post-infall processes, where galaxies entering dense environments undergo accelerated quenching. This disparity is primarily driven by environmental mechanisms such as ram-pressure stripping, which removes interstellar gas through interactions with the intracluster medium \citep{gunn1972infall}. Also, starvation (strangulation) or harassment could effectively decrease gas in these galaxies.
In contrast, void galaxies may follow a different quenching pathway, in which gas mass depletion is possibly driven by stellar feedback -- including radiation pressure, stellar winds, and supernova-driven outflows -- rather than by environmental mechanisms \citep{dalla2008simulating}. We note that this interpretation is not directly tested by our analysis and is offered here as a plausible explanation consistent with the absence of environmental stripping in voids, rather than as an established conclusion.

In Figure \ref{fig:01stellar_mass_comparison03} the evolution of stellar mass in our samples indicates that quenched galaxies evolve more rapidly than star-forming galaxies, regardless of their environments. Interestingly, quenched void galaxies exhibit higher host halo dark matter masses over the past 6 Gyr ($z < 0.6$) of evolution. This trend is particularly significant in larger stellar mass bins. More dark matter by quenched void galaxies may be efficient in accelerating the star formation efficiency (the fraction of gas converted into stars on a free-fall time) and lead to rapid gas consumption,
that are more effective at higher redshifts \citep{boylan2024accelerated}.
In contrast, quenched cluster galaxies do not exhibit a positive correlation with redshift over the last 8 Gyr; instead, they maintain a relatively constant dark matter mass throughout their evolution. Additionally, less massive dark matter content in galaxies within clusters may impact the global tidal field of the cluster potential well, which could be strong enough to truncate the dark matter halos of galaxies \citep{limousin2007truncation}.  

For star-forming and quenched galaxies, a galaxy is defined as gas-rich if its gas fraction fulfills the criterion: 
\(f_{\rm gas} > 0.5\)  \citep{stewart2009gas}. 
As Figure~\ref{fig:01stellar_mass_comparison03} illustrates, in the lower stellar mass range \(\left[10^{8.5} - 10^{9.5}\right]\, M_\odot\), our star-forming galaxies in both environments, as well as quenched galaxies in voids, remain gas-rich across all redshifts. However, across all stellar mass bins, quenched galaxies in clusters exhibit a steeper decline in gas fraction compared to other populations. Furthermore, galaxies at higher redshifts exhibit significantly higher gas fractions than their low-redshift counterparts, consistent with theoretical expectations that earlier cosmic epochs were characterized by abundant gas supplies facilitating active star formation \citep{tacconi2020evolution}. The distinct evolutionary behavior of gas fraction in quenched galaxies across environments indicates that the environment exerts an additional effect on the gas content of quenched galaxies, which is evident in clusters but not observed in voids.

\begin{figure*}[t] 
    \centering
    \includegraphics[width=0.9\textwidth, keepaspectratio]{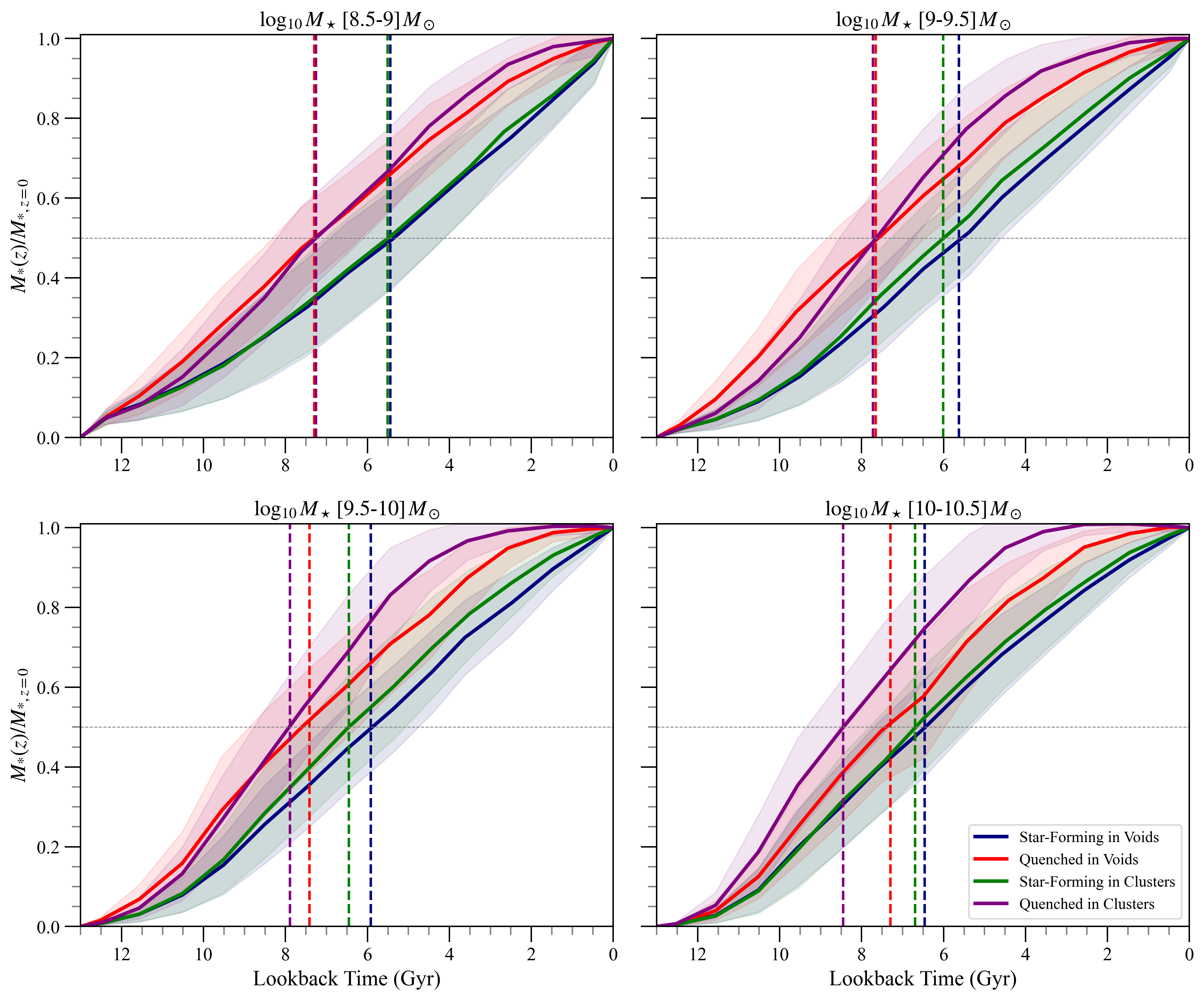}
 \caption{Stellar mass assembly histories of star-forming (blue) and quenched (red) galaxies in voids, and star-forming (green) and quenched (purple) galaxies in clusters, shown across four stellar mass bins as a function of lookback time (Gyr). The normalized stellar mass fraction is plotted, with solid lines representing the median trends and shaded regions indicating the median absolute deviation (MAD). Vertical dashed lines mark the formation time ($t_{\text{form}}$), defined as the time when 50\% of the final stellar mass is assembled. Gray horizontal dashed lines indicate the 50\% assembly thresholds of the final stellar mass.}

    \label{fig:stellar_mass_6comparison}
\end{figure*}
\subsection{Stellar Mass Assembly Histories of Void and Cluster Galaxies}

This section analyses the evolution of star-forming and quenched galaxies in voids and clusters by examining their stellar mass assembly histories. To quantify the formation time of galaxies, we define \(t_{\text{form}}\) as the lookback time at which the main progenitor has assembled 50\% of its final stellar mass at \(z=0\). To calculate \(t_{\text{form}}\), we use a normalized stellar mass fraction defined as \(M_*(z)/M_{*, z=0}\), where \(M_*(z)\) is the stellar mass at a given redshift.

Figure~\ref{fig:stellar_mass_6comparison} shows the normalized stellar mass
assembly histories, \(M_\star(z)/M_{\star,z=0}\), for star-forming and quenchedgalaxies in voids and clusters across four stellar-mass bins. In each lookback-time bin, we plot the median normalized mass fraction to trace the typical stellar mass growth of each population, while the shaded regions show the median absolute deviation (MAD) about the median, providing a robust measure of the scatter in the assembly histories. The vertical dashed lines mark \(t_{\mathrm{form},50\%}\), the lookback time at which each population has assembled 50\% of its final stellar mass. Blue and green lines correspond to star-forming galaxies in voids and clusters, respectively, while red and purple lines indicate their quenched counterparts. The median values of
\(t_{\mathrm{form},50\%}\) and their 25th--75th percentile ranges are listed in Table~\ref{tab:tform50_samples}.

Two systematic trends are evident. First, quenched galaxies assemble
their stellar mass earlier than star-forming galaxies, largely independent
of large-scale environment. In every stellar-mass bin, the median assembly
curves of quenched galaxies are shifted toward earlier lookback times
relative to those of their star-forming counterparts, indicating that
quenched systems had already formed a larger fraction of their present-day
stellar mass at earlier epochs. This behaviour is consistent with a
scenario in which quenching is associated with an earlier phase of rapid
stellar mass growth, followed by reduced cold-gas availability and
suppressed star formation at later times \citep{peng2010mass,
behroozi2013sfh, wechsler2018halos}.

Second, the role of environment depends on both stellar mass and
evolutionary stage. For formation times, the void--cluster separation
among quenched galaxies is small at low stellar masses
($\log_{10} M_\star/\mathrm{M_\odot} \lesssim 9.5$), where quenched
galaxies in voids and clusters reach their 50 per cent formation time at
nearly the same lookback time. This environmental offset grows with
stellar mass, becoming most pronounced at $\log_{10}
M_\star/\mathrm{M_\odot} \sim 10$--$10.5$, where quenched cluster
galaxies form earlier than their void counterparts. For star-forming
galaxies, the environmental difference in formation time remains weak
across all stellar-mass bins.

The most distinct environmental signature appears at late times, after
galaxies have assembled more than half of their final stellar mass. At
this stage, quenched cluster galaxies have already assembled the large
majority of their stellar mass and their assembly curves flatten toward
$z=0$, whereas quenched void galaxies continue to assemble a significant
fraction of their stellar mass, showing more extended residual growth.
This contrast is visible across the quenched population but strengthens
substantially toward higher masses: at $\log_{10}
M_\star/\mathrm{M_\odot} \sim 9.5$--$10.5$, quenched cluster galaxies
approach their final stellar mass earlier, while their void counterparts
continue to grow more gradually. Star-forming galaxies, by contrast,
continue to assemble their stellar mass actively at late times
regardless of environment, with at most a weak void--cluster offset in
assembly rate at the highest stellar masses.

This behaviour is consistent with previous studies showing that galaxy
evolution proceeds more slowly in underdense environments
\citep{dominguez2023galaxies, rodriguez2024evolutionary}. Our results
extend this picture by showing that the environmental dependence of
late-time assembly is present across all stellar masses in the quenched
population, growing in strength toward higher masses, while star-forming
galaxies are only weakly affected by their large-scale surroundings.

\renewcommand{\arraystretch}{1.5}
\begin{table}[h!]
\centering
\begin{tabular}{|c|c|c|c|c|}
\hline
\textbf{Mass bins} & \multicolumn{2}{c|}{\textbf{Star-Forming (Gyr)}} & \multicolumn{2}{c|}{\textbf{Quenched (Gyr)}} \\
\hline
\textbf{$\log_{10}(M_\star)$} & \textbf{Voids} & \textbf{Clusters} & \textbf{Voids} & \textbf{Clusters} \\
\hline
\textbf{[8.5--9]}
& $5.44 \, \substack{-1.30 \\ +1.40}$
& $5.52 \, \substack{-1.50 \\ +1.41}$
& $7.30 \, \substack{-0.99 \\ +0.65}$
& $7.26 \, \substack{-1.15 \\ +0.91}$ \\
\textbf{[9--9.5]}
& $5.63 \, \substack{-1.17 \\ +1.13}$
& $6.01 \, \substack{-1.36 \\ +1.24}$
& $7.66 \, \substack{-1.19 \\ +1.21}$
& $7.73 \, \substack{-0.99 \\ +0.74}$ \\
\textbf{[9.5--10]}
& $5.92 \, \substack{-1.09 \\ +1.12}$
& $6.46 \, \substack{-1.29 \\ +1.22}$
& $7.42 \, \substack{-0.85 \\ +1.91}$
& $7.89 \, \substack{-0.94 \\ +0.92}$ \\
\textbf{[10--10.5]}
& $6.46 \, \substack{-1.20 \\ +1.18}$
& $6.70 \, \substack{-1.31 \\ +1.20}$
& $7.30 \, \substack{-1.15 \\ +1.54}$
& $8.46 \, \substack{-0.89 \\ +0.90}$ \\
\hline
\end{tabular}

\caption{Median lookback times ($t_{\mathrm{form},50\%}$) for star-forming and quenched galaxies in void and cluster environments, defined as the time at which galaxies have assembled 50\% of their final stellar mass at $z=0$. The subscript and superscript values indicate the differences between the median and the 25th and 75th percentiles, respectively. The number of galaxies in each sample is reported in the text.}
\label{tab:tform50_samples}
\end{table}

\begin{figure}[t] 
    \centering
    \includegraphics[width=0.5\textwidth]{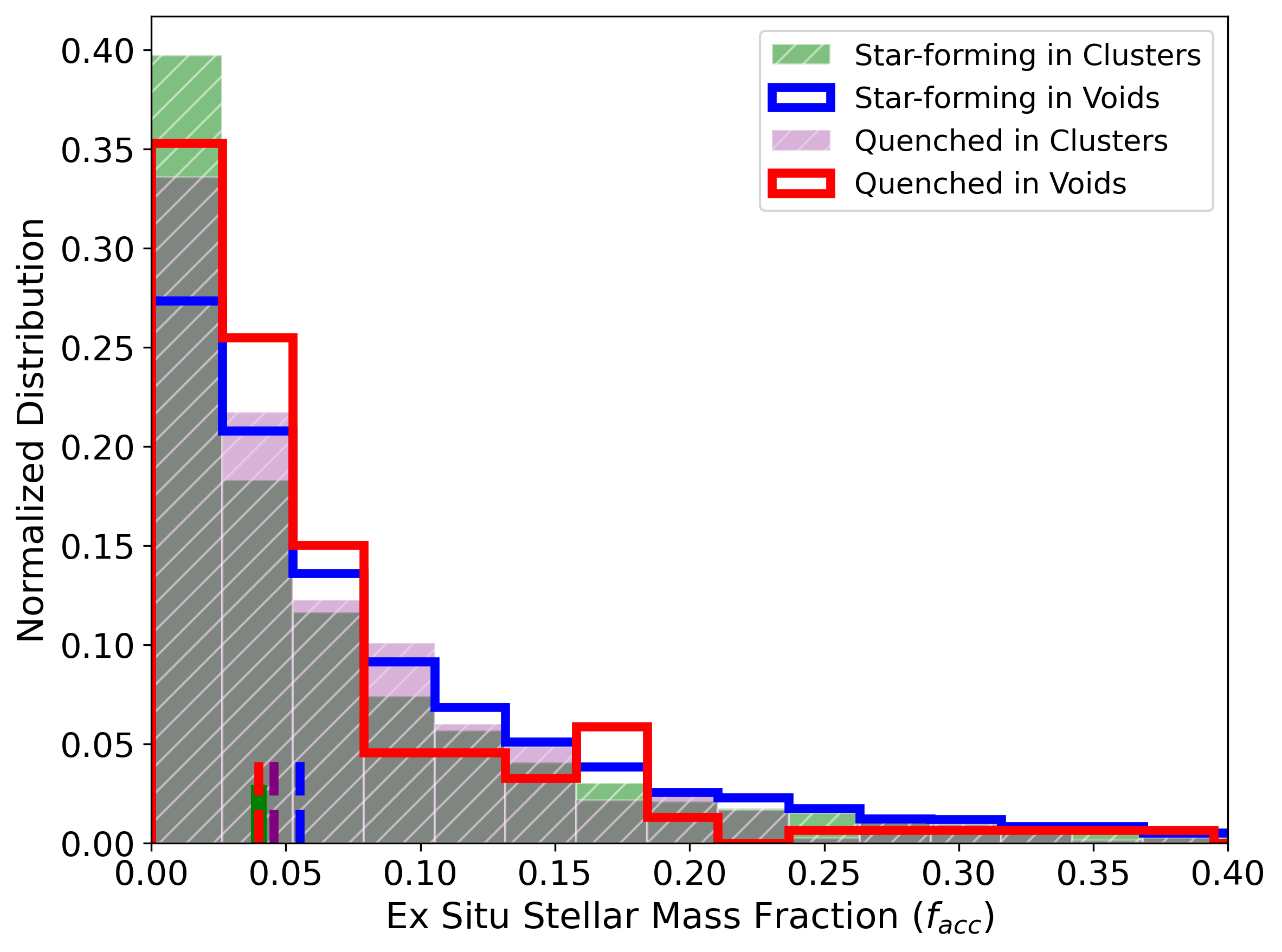}
 \caption{The normalized histogram of the ex-situ stellar mass fraction (\( f_{\text{acc}} \)) across different galaxy populations, categorized by environment (voids vs. clusters) and star-forming versus quenched galaxies. The small vertical lines indicate the corresponding median values for each population.}

    \label{fig:ex-situ_stellar_mass} 
\end{figure}

\subsection{Merger Histories of Void and Cluster Galaxies}
Understanding the importance of mergers in the context of galaxy formation history requires investigating the rates of galaxy mergers over various cosmic periods \citep{conselice2009structures}. As voids are environments with a low density of galaxies, the rate of galaxy interactions is expected to be lower than in cluster environments \citep{van2011cosmic}. However, simulations predict that the number of galaxy mergers does not depend on the environment, and galaxies within voids have undergone more recent mergers compared to galaxies in other environments, indicating a different assembly rate \citep{rosas2022revealing, rodriguez2024evolutionary}. 
For cluster galaxies, it was often cited that the velocity dispersion is too high to allow mergers. This is a result of the galaxies approaching with too high a relative velocity to coalesce. This is despite the higher density that results in a higher incidence of interactions \citep{ostriker1980elliptical, mihos2004interactions, oh2018kydisc}. Clusters, therefore, are typically seen to have fewer mergers than the field environment \citep[e.g.][]{tonnesen2012effects, delahaye2017galaxy}. The outer regions of clusters, where the velocity dispersion is potentially lower, can have galaxies approaching slow enough to result in mergers \citep{mihos2004interactions, deger2018tidal, kelkar2019time}. Additionally, galaxy interactions are particularly effective at stimulating substantial star formation in low- to moderate-density environments \citep{sol2006effects, das2021galaxy}. Furthermore, wet mergers are more prevalent in under-dense regions \citep{lin2010wet}. These studies inspired us to undertake a statistical analysis of merger rates in void and cluster environments, focusing on star-forming and quenched low-mass galaxies. 

For our purpose, we use the publicly available merger history catalogue of \citet{rodriguez2017role} and \citet{eisert2023ergo}, which contains information regarding galaxy mergers and the ex-situ (accretion of stars, where stars are born in other smaller galaxies) and in-situ (stars are formed within the galaxy from infalling cold gas) stellar formation processes. The merger time is defined as the time the corresponding branches of the merger tree join. Additionally, for a deep analysis of the effect of mergers on star-forming and quenched galaxies, mergers are categorized into three types: major (stellar mass ratio > 1/4), minor (stellar mass ratio between 1/10 and 1/4), and mini (stellar mass ratio < 1/10) mergers. The mass ratio is always based on the stellar masses of the two merging galaxies when the secondary reaches its maximum stellar mass \citep{rodriguez2015merger}. 

To ensure a fair comparison of merger statistics between void and cluster galaxies, we mass-matched the star-forming and quenched samples across environments. Galaxies were first restricted to the stellar-mass range $8.5 \leq \log_{10}(M_\star/M_\odot) \leq 10.5$. We then divided the samples into 0.1 dex stellar-mass bins and randomly down-sampled each population in every bin to match the minimum normalized fraction among the four samples.  After mass matching, the final samples consist of 9,056 star-forming and 153 quenched galaxies in voids, and 7,929 star-forming and 783 quenched galaxies in clusters. This ensures that the stellar-mass distributions are consistent across all populations, allowing for an unbiased comparison of merger rates and statistics.

The primary factor we use to quantify the importance of merging history is the \textit{ex-situ stellar mass fraction} (the fraction of the total stellar mass of a galaxy that is contributed by stars that formed in other galaxies and were subsequently accreted as a consequence of the hierarchical growth of structures) symbolized by \( f_{\text{acc}} \). It is essential to emphasize that \( f_{\text{acc}} \) does not directly measure the history of merging; instead, it evaluates the significance of dry merging concerning dissipative processes, such as in-situ star formation \citep{oser2010two}. Figure \ref{fig:ex-situ_stellar_mass} presents normalized distributions of \textit{\( f_{\text{acc}} \)} over the last 12.6 Gyr ( $z<5$) for our samples of star-forming and quenched galaxies within both cluster and void environments. The corresponding data were derived from the merger history catalogues within the TNG300 simulation \citep{rodriguez2017role, eisert2023ergo}. The small vertical lines indicate the corresponding median values for each population.

As discussed in \citet{rodriguez2017role}, at any fixed stellar mass, $f_{\rm acc}$ is negatively correlated with gas-rich mergers, and higher $f_{\rm acc}$ is associated with a more significant number of massive, recent, and dry mergers that are particularly significant for massive galaxies ($M_* > 10^{11}\,M_\odot$). The observed median values of \textit{ex-situ stellar mass fraction}, remaining below 0.1 across our void and cluster samples, indicate that the stellar mass growth of galaxies is dominated by \textit{in-situ stellar assembly} at all redshifts. These results indicate that mergers between our low-mass galaxies tend to be gas-rich throughout their histories.

To assess whether the modest differences visible in Figure~\ref{fig:ex-situ_stellar_mass} are statistically meaningful, we performed pairwise Kolmogorov--Smirnov (K--S) tests on the underlying $f_{\rm acc}$ distributions for all four populations. Five of the six pairwise comparisons yield statistically significant differences ($p < 0.005$); the only exception is the comparison between quenched galaxies in voids and in clusters ($p = 0.29$), which are statistically indistinguishable in $f_{\rm acc}$. Notably, star-forming galaxies in voids exhibit the highest median value (median $f_{\rm acc} = 0.055$), significantly higher than all other populations, whereas quenched galaxies show no significant environmental dependence in this quantity. This indicates that, in contrast to several of the properties discussed in Sect.~3.2, the ex-situ mass fraction of quenched galaxies is not strongly shaped by large-scale environment, while star-forming galaxies in voids stand out as accumulating a modestly higher fraction of their stellar mass through mergers.

\begin{figure*}[t] 
    \centering
    \includegraphics[width=\textwidth, keepaspectratio]{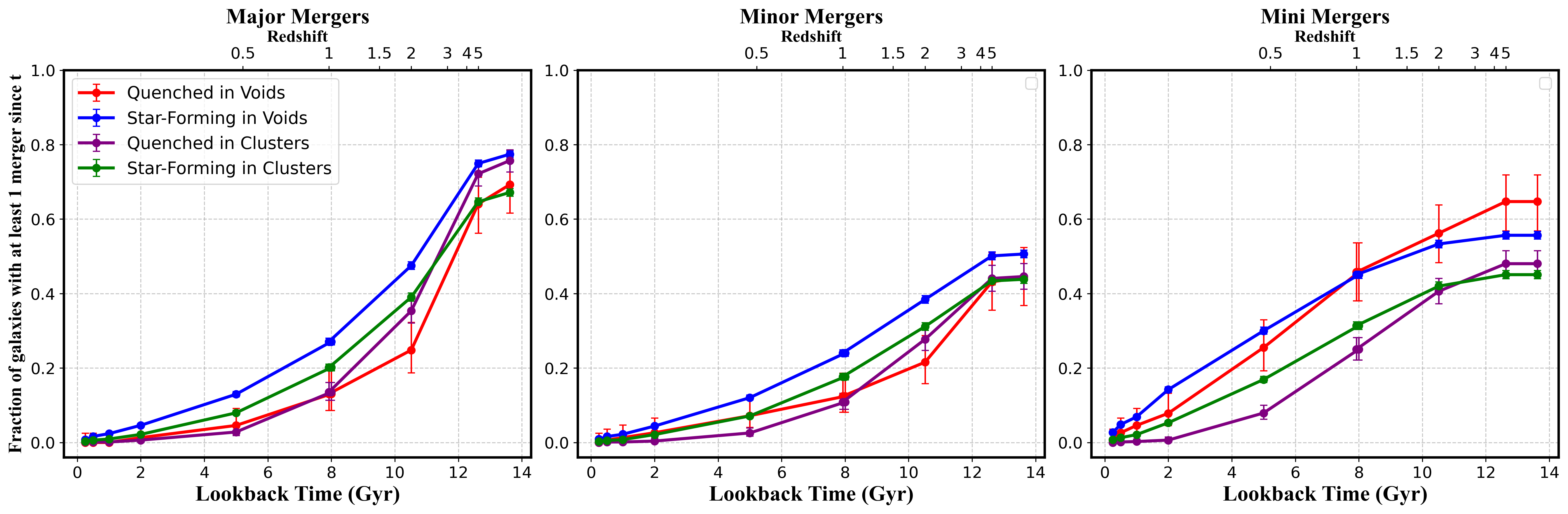}
 \caption{Cumulative fraction of galaxies that have experienced at least one merger since a given lookback time, for different stellar mass ratios corresponding to major, minor, and mini mergers for stellar-mass–matched samples. By construction, the value at each lookback time $t$ includes all mergers that occurred at any time between $t$ and $z=0$, rather than mergers occurring specifically at $t$; these curves therefore represent integrated merger histories rather than instantaneous merger rates. Blue and red lines show star-forming and quenched galaxies in voids, respectively, while green and purple lines correspond to star-forming and quenched galaxies in clusters. Dots indicate the discrete lookback times at which the measurements are evaluated, and error bars show the 95\% Wilson-score binomial confidence interval at each point, reflecting the sample size of each population.}
    \label{fig:stellar_mass_8comparison}
\end{figure*}

\subsubsection{Merger statistics}
Figure~\ref{fig:stellar_mass_8comparison} shows the cumulative fraction of star-forming and quenched galaxies in voids and clusters that have experienced at least one merger as a function of lookback time, for major, minor, and mini mergers. By construction, the cumulative merger fraction increases monotonically toward higher lookback times, reflecting the integrated merger history of galaxies rather than the instantaneous merger rate. Thus, this figure does not directly trace the redshift evolution of the merger rate, but instead provides a comparative view of the merger histories across different galaxy populations and environments. In this context, differences between the curves indicate variations in the likelihood and timing of merger events rather than changes in merger rates.

Our results show that star-forming galaxies in voids exhibit a systematically higher cumulative merger fraction (defined as the fraction of galaxies that have experienced at least one merger since a given lookback time) compared to other populations, particularly for both major and minor mergers. In addition, void galaxies demonstrate a notably higher fraction of mini mergers than their cluster counterparts, suggesting that low-mass accretion events are more prevalent in low-density environments.

\begin{figure*}[t] 
    \centering
    \includegraphics[width=\textwidth, keepaspectratio]{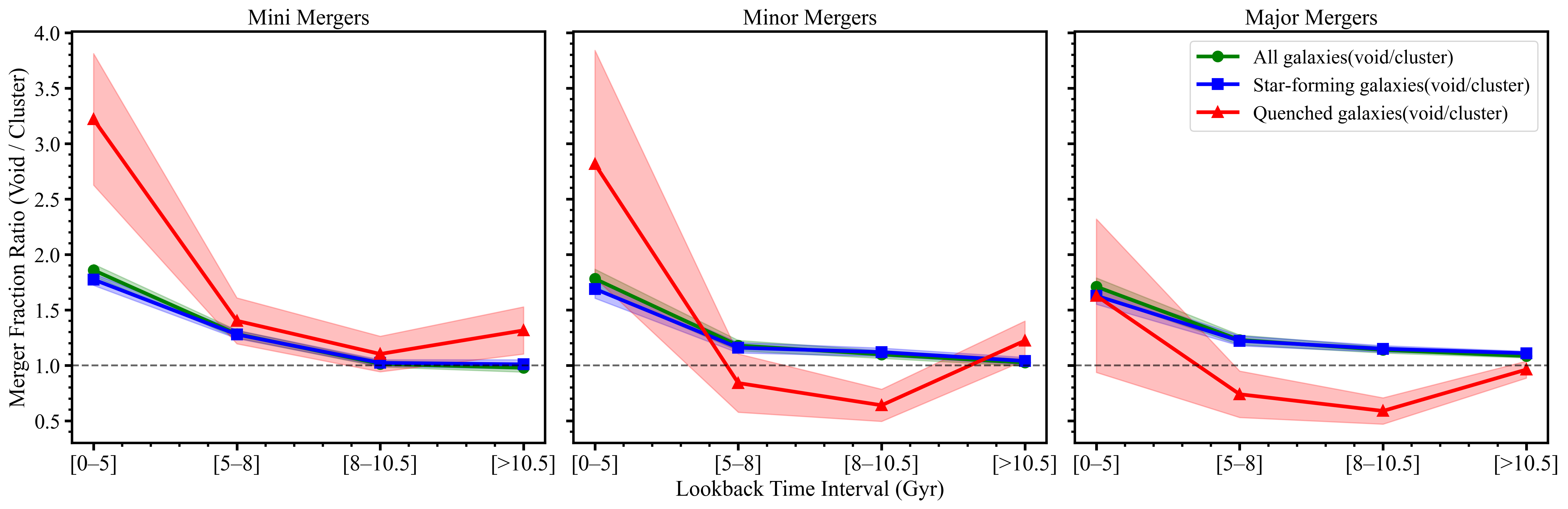}
\caption{Void-to-cluster merger-fraction ratio, $R$, as a function of lookback time for mini (left), minor (middle), and major (right) mergers for stellar-mass–matched samples, shown separately for all (green), star-forming (blue), and quenched (red) galaxies. The four lookback-time bins correspond to [0--5], [5--8], [8--10.5], and [$>10.5$] Gyr, and each bin represents the merger fraction accumulated \emph{strictly within} that discrete interval, in contrast to the cumulative-since-$z=0$ fractions shown in Fig.~\ref{fig:stellar_mass_8comparison}. The ratio is defined as the merger fraction in void galaxies divided by that in cluster galaxies. Values of $R>1$ indicate higher merger fractions in voids, while $R<1$ indicates higher merger fractions in clusters. The shaded regions show the propagated $1\sigma$ uncertainties, computed from binomial errors on the merger fractions in each environment. Markers denote the central (best-estimate) value of $R$ in each bin.}

    \label{fig:merger_ratio}
\end{figure*}

We emphasize that, by construction, the cumulative fraction at a given lookback time $t$ already includes all mergers that occurred at any point between $t$ and $z=0$; it therefore reflects the full accumulated merger history since $t$, not merger activity occurring specifically at that epoch. With this in mind, we note that by the earliest lookback times sampled ($\sim 12$--14 Gyr), cumulative fractions already differ between populations: star-forming galaxies in voids show a higher cumulative fraction of major mergers than quenched galaxies in voids, whereas in clusters, quenched and star-forming galaxies show comparable cumulative fractions. Moving to intermediate lookback times (8--12 Gyr), these differences become more pronounced, with quenched cluster galaxies reaching higher cumulative major-merger fractions than quenched void galaxies. Over the last $\sim 10.5$ Gyr, star-forming galaxies consistently reach higher cumulative fractions than quenched galaxies in both environments, with the difference being more significant in voids. A similar pattern holds for minor mergers, with star-forming galaxies in both environments reaching higher cumulative fractions than quenched galaxies over the last $\sim 12$ Gyr. The differences in the cumulative fraction of mini mergers between void and cluster galaxies further highlight the role of environment, with void galaxies more likely to experience low-mass mergers, reflecting the distinct dynamical conditions in under-dense regions. We note that the quenched samples (N=153 in voids, N=783 in clusters) are considerably smaller than the star-forming samples (N=9,056 and N=7,929, respectively), and the associated binomial uncertainties, shown as error bars in Fig.~\ref{fig:stellar_mass_8comparison}, are correspondingly larger; differences involving quenched populations, particularly at intermediate lookback times, should be interpreted with this in mind. These results collectively indicate that merger histories depend both on environment and star formation activity, with star-forming galaxies generally experiencing a larger number of merger events over their evolutionary history.

Figure~\ref{fig:merger_ratio} presents the void-to-cluster merger-fraction ratio, $R$, for mini, minor, and major mergers, separated into all, star-forming, and quenched galaxy populations, as a function of lookback time. The analysis is performed in four intervals: [0--5], [5--8], [8--10.5], and [$>10.5$] Gyr. Values above unity indicate higher merger fractions in void galaxies, while values below unity indicate enhanced merger activity in cluster environments. The shaded regions show the propagated $1\sigma$ uncertainties derived from binomial errors, which remain relatively small for the total and star-forming samples but increase significantly for quenched galaxies due to limited number statistics. Overall, the figure reveals a clear environment-driven and time-dependent evolution in merger activity. At recent times ([0--5] Gyr), the merger-fraction ratio is systematically above unity for nearly all merger types and populations, indicating that void galaxies experience enhanced merger activity compared to cluster galaxies. Toward earlier epochs, the ratios approach unity, showing that environmental differences weaken at higher lookback times.

For the total galaxy population in each environment, all merger types show enhanced ratios at recent times, with $R \sim 1.6$--$1.8$, while at earlier epochs the ratios decrease toward $R \sim 1.0$--$1.2$. This indicates that, globally, merger activity in voids increases toward the present day, whereas at earlier cosmic times the merger histories of void and cluster galaxies are broadly similar. The star-forming population exhibits a smooth and consistent trend across all merger types. The ratio gradually increases toward the present day, reaching $R \sim 1.6$--$1.8$ at [0--5] Gyr, and declines to $R \sim 1.0$--$1.2$ at earlier times. This indicates that star-forming galaxies in voids systematically experience more mergers than those in clusters at recent times, although the environmental dependence remains moderate. In contrast, the quenched population shows the strongest and most pronounced environmental dependence. At recent times, quenched galaxies in voids exhibit a dramatic enhancement in merger activity, particularly for low-mass-ratio events. Mini mergers reach $R \sim 3.0$--$3.3$, and minor mergers reach $R \sim 2.7$--$2.9$, indicating that quenched galaxies in voids experience significantly more small mergers at late times than quenched galaxies in clusters. This suggests that low-mass accretion plays a key role in the continued evolution of quenched systems in low-density environments.

At intermediate lookback times ([5--8] and [8--10.5] Gyr), a clear reversal is observed for quenched galaxies. For minor and major mergers, the ratios drop below unity, reaching $R \sim 0.6$--$0.8$, indicating that quenched galaxies in cluster environments experienced more merger activity than those in voids during these epochs. This likely reflects enhanced merger rates during group and cluster assembly in dense environments. For mini mergers, the quenched ratio decreases significantly from $R \sim 3.2$ at recent times to $R \sim 1.3$ at [5--8] Gyr and $R \sim 1.1$ at [8--10.5] Gyr. Although it remains above unity, the environmental contrast weakens, indicating that small mergers are consistently more frequent in voids but become less environment-dependent at earlier times.

Taken together, these results show that environment strongly shapes merger activity , with its impact depending on galaxy type, merger mass ratio, and cosmic time. At recent epochs, void galaxies exhibit enhanced merger activity across mini, minor, and major mergers, with the strongest increase in quenched systems. In contrast, at intermediate times, quenched cluster galaxies show the highest merger activity—especially for major and minor mergers. Star-forming galaxies follow a smoother trend: merger activity in voids increases steadily toward the present day but never shows the strong reversals seen in quenched systems.

Research into the effects of merger mass ratios on star formation and gas consumption reveals remarkable insights. \citet{bottrell2024illustristng} found that mini mergers lead to modest yet statistically significant enhancements in star formation rates (SFRs) compared to non-merging galaxies and have a substantial role in in-situ stellar mass assembly in galaxies. Their study also showed that while major mergers are consistent after just 1 Gyr (short-lived), on average, mini mergers are enhanced in SFR for as long as 3 Gyr (long-lived).  But our findings open a unique new window into this perspective, revealing that mini mergers occur far more frequently in under-dense environments (voids) compared to dense clusters, regardless of whether the galaxies are star-forming or quenched, within stellar mass ranges \( 10^{8.5} - 10^{10.5} \, M_{\odot} \). A detailed breakdown of the merger fractions—including the relative contributions of major, minor, mini, and multiple mergers in both void and cluster environments—is provided in Appendix~\ref{appendix:Additional Merger Statistics} (Figure~\ref{fig:stellar_mass_080comparison}).

\begin{figure*}[t]
    \centering
    
    \includegraphics[width=\textwidth, keepaspectratio]{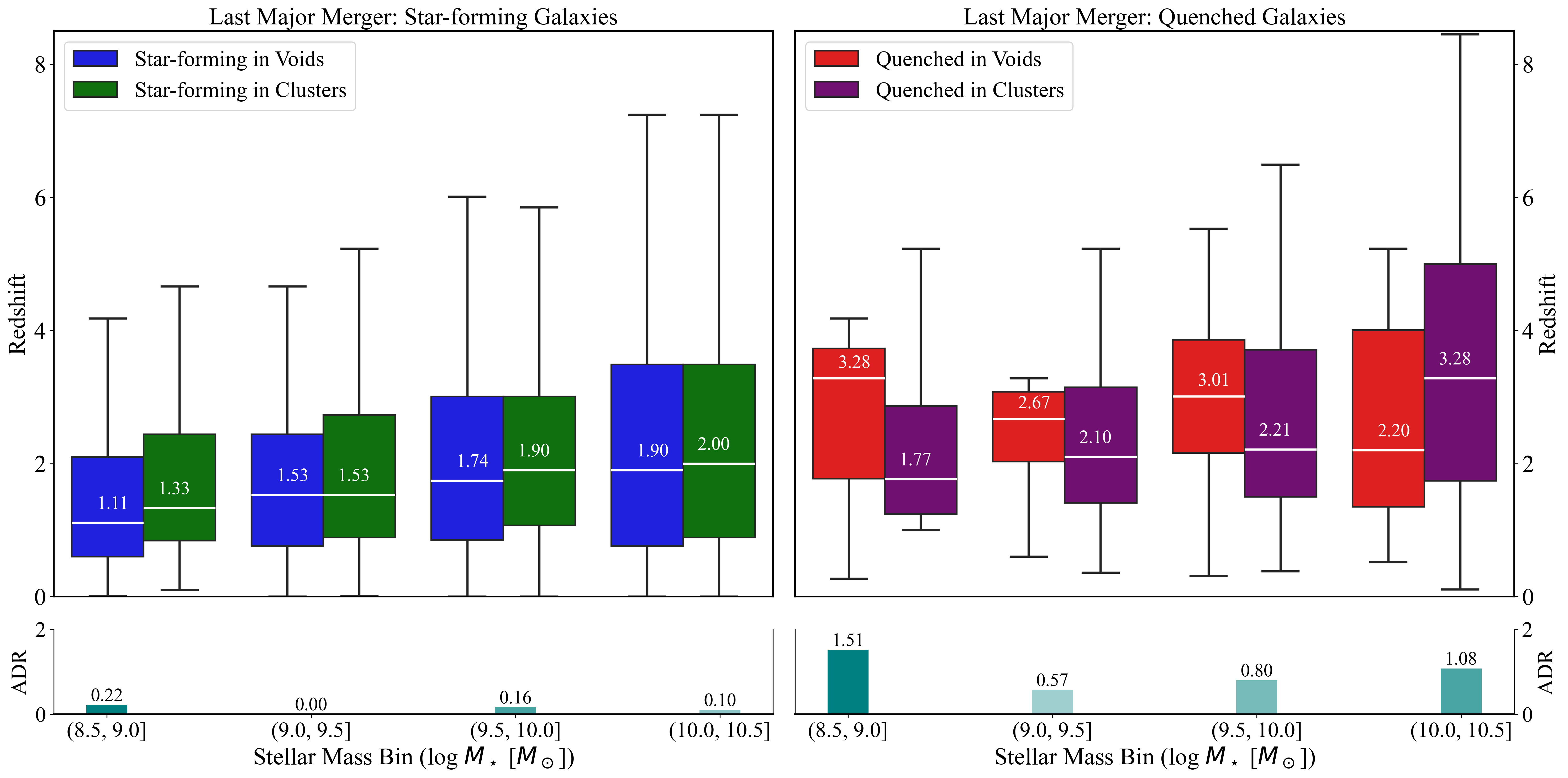}
    
    \vspace{0.5em} 
    
    \includegraphics[width=\textwidth, keepaspectratio]{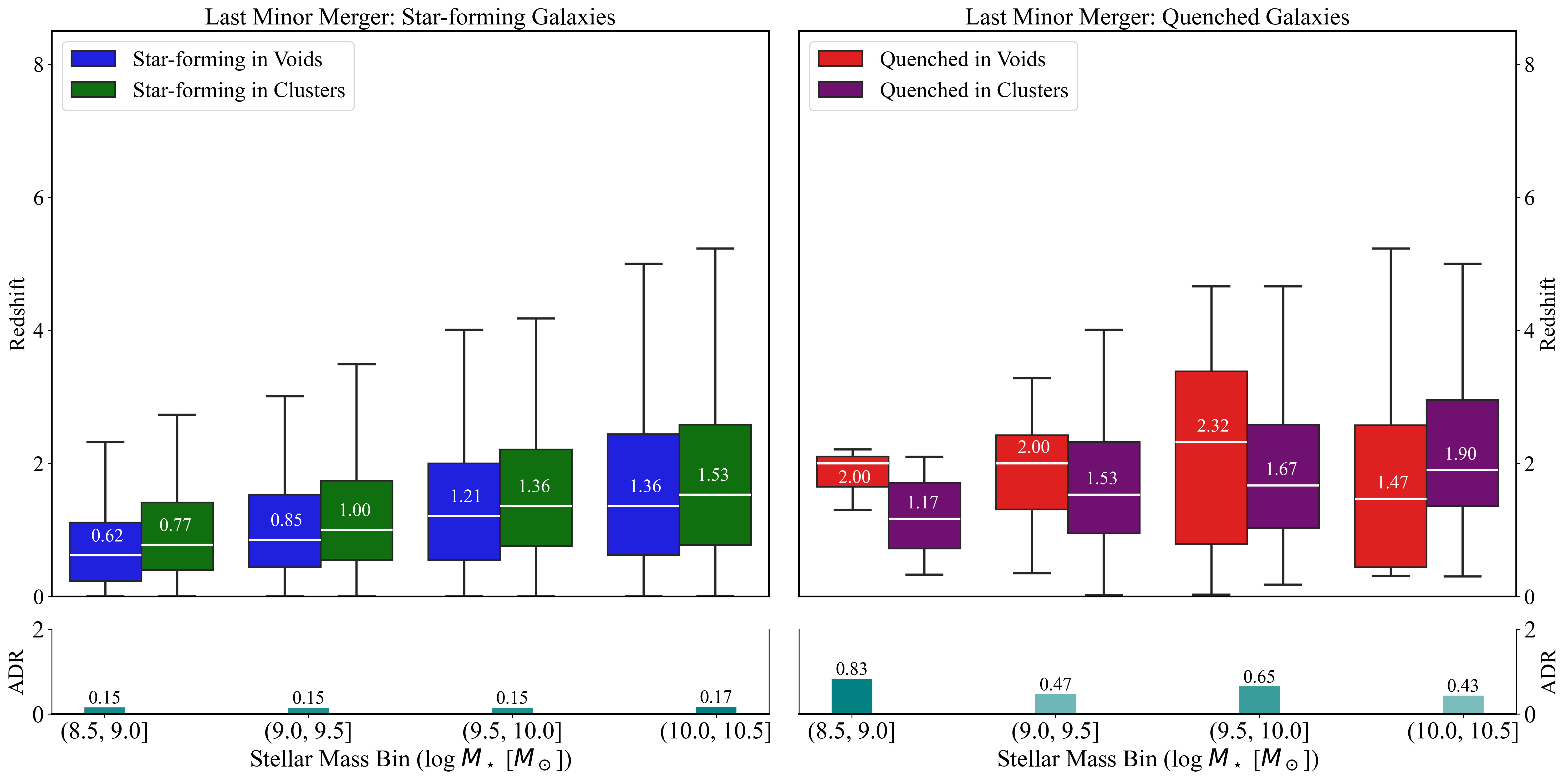}
\caption{Comparison of the redshift of the last major (top) and minor (bottom) mergers for star-forming (left) and quenched (right) galaxies, comparing void and cluster environments directly within each galaxy type, across four stellar mass bins. Bar charts below each panel show the absolute difference in redshift (ADR) between void and cluster environments for the corresponding galaxy type, merger class, and mass bin. Each horizontal white line in the box plots represents the median redshift of the last merger in each stellar mass bin.}
    \label{fig:01stellar_mass_comparison010}
\end{figure*}

\begin{figure}[t] 
    \centering
    \includegraphics[width=0.5\textwidth]{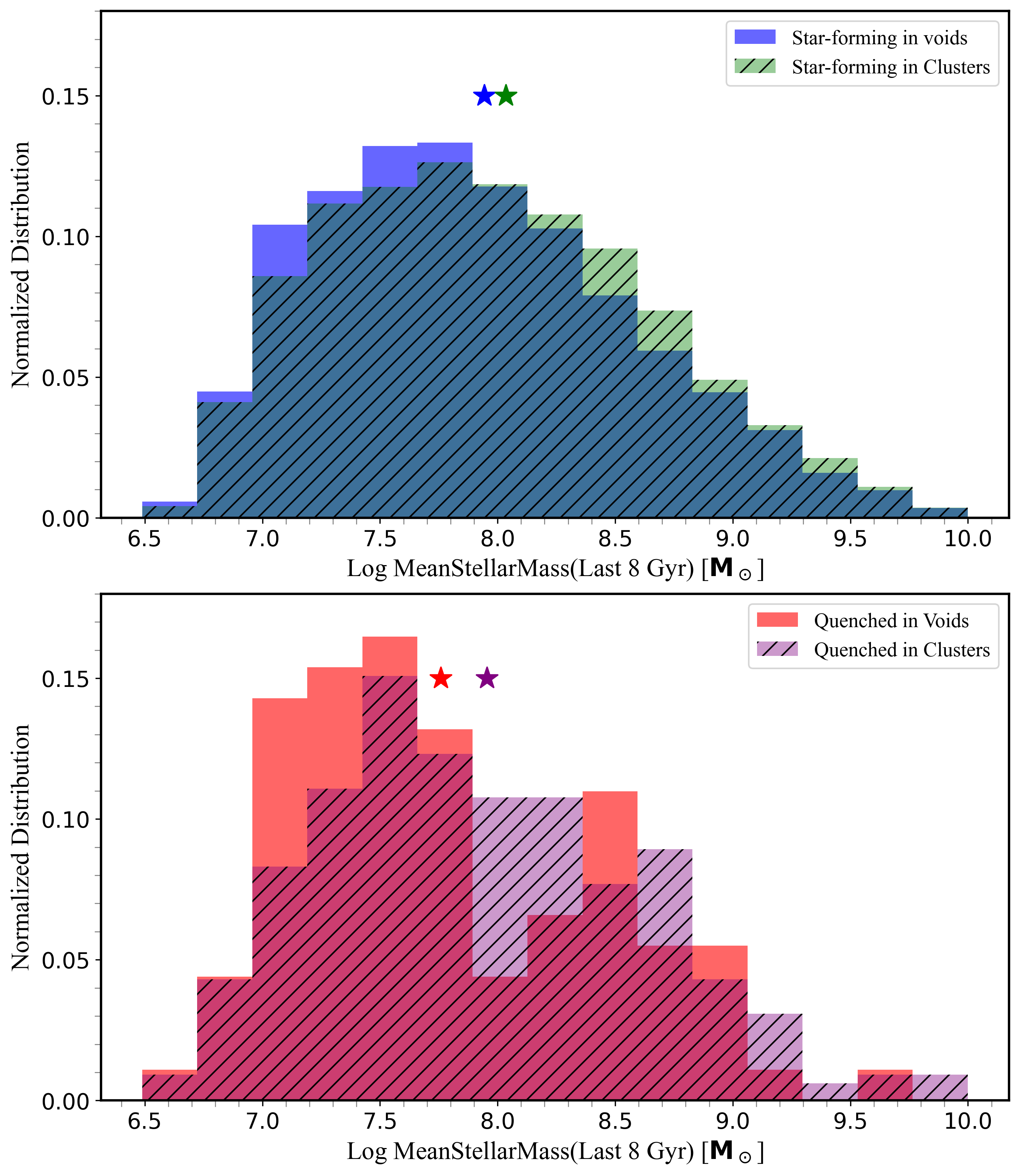}
\caption{
Normalized histogram of the log$_{10}$ mean stellar mass ($M_\odot$) over the last 8~Gyr for the different galaxy populations, 
categorized by environment (voids vs. clusters) and galaxy type (star-forming vs. quenched). 
Stars indicate the corresponding median values for each population. 
All distributions are shown for stellar-mass–matched samples, ensuring that differences between environments and galaxy types 
are not driven by underlying differences in stellar-mass distributions.
}
    \label{fig:mean_stellar_mass_ratio} 
\end{figure}
\begin{table}[h!]
\centering
\caption{Sample sizes (non-merger / merger, $N_{\rm mergers} \geq 1$, over the last 8 Gyr) for the subsamples of star-forming (SF) and quenched galaxies used in the merger-versus-non-merger comparison of Sect.~3.6, separated by environment and stellar-mass bin. Low-mass and high-mass correspond to $10^{8.5}$--$10^{9.5}\,M_\odot$ and $10^{9.5}$--$10^{10.5}\,M_\odot$, respectively.}
\begin{tabular}{lcc}
\hline
 & \textbf{Voids} & \textbf{Clusters} \\ \hline
Quenched, low-mass  & 40 / 46     & 298 / 144  \\
Quenched, high-mass & 22 / 45     & 160 / 181  \\
SF, low-mass        & 1960 / 3173 & 2615 / 1872 \\
SF, high-mass       & 729 / 3194  & 1156 / 2286 \\ \hline
\end{tabular}
\label{tab:merger_nonmerger_samples}
\end{table}

\begin{figure*}[!t] 
    \centering
    
    \includegraphics[width=0.9\textwidth, keepaspectratio]{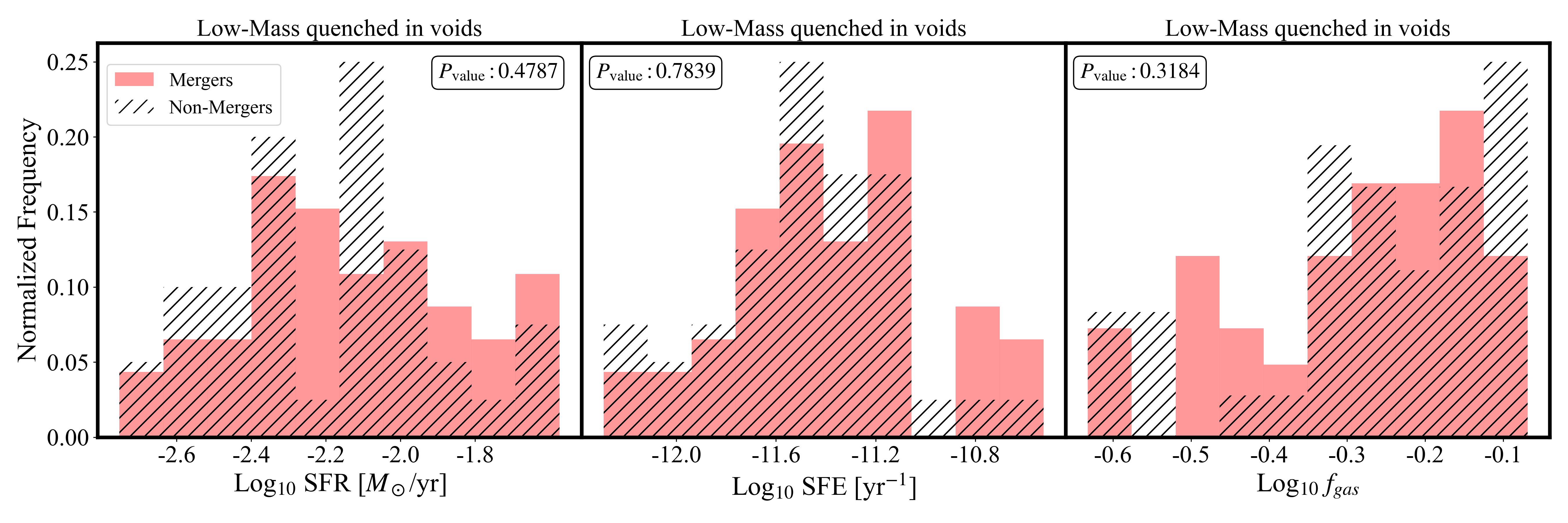}
    
    \vspace{0.1em} 
    
    \includegraphics[width=0.9\textwidth, keepaspectratio]{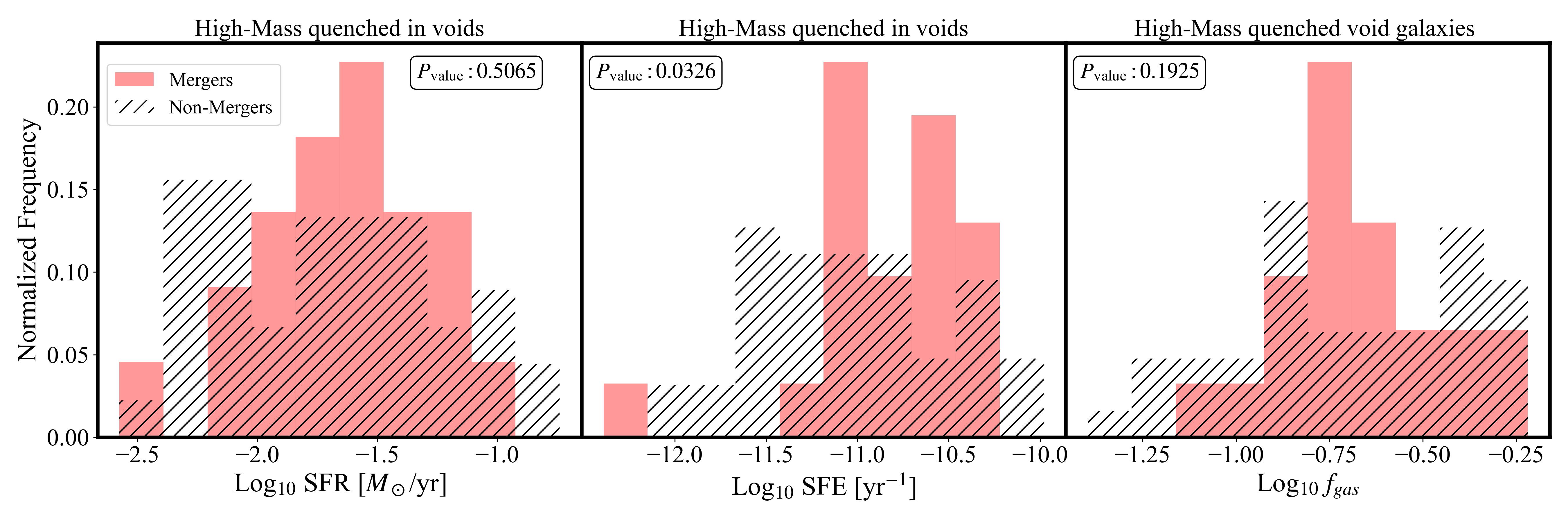}
    
    \vspace{0.1em} 
    
    \includegraphics[width=0.9\textwidth, keepaspectratio]{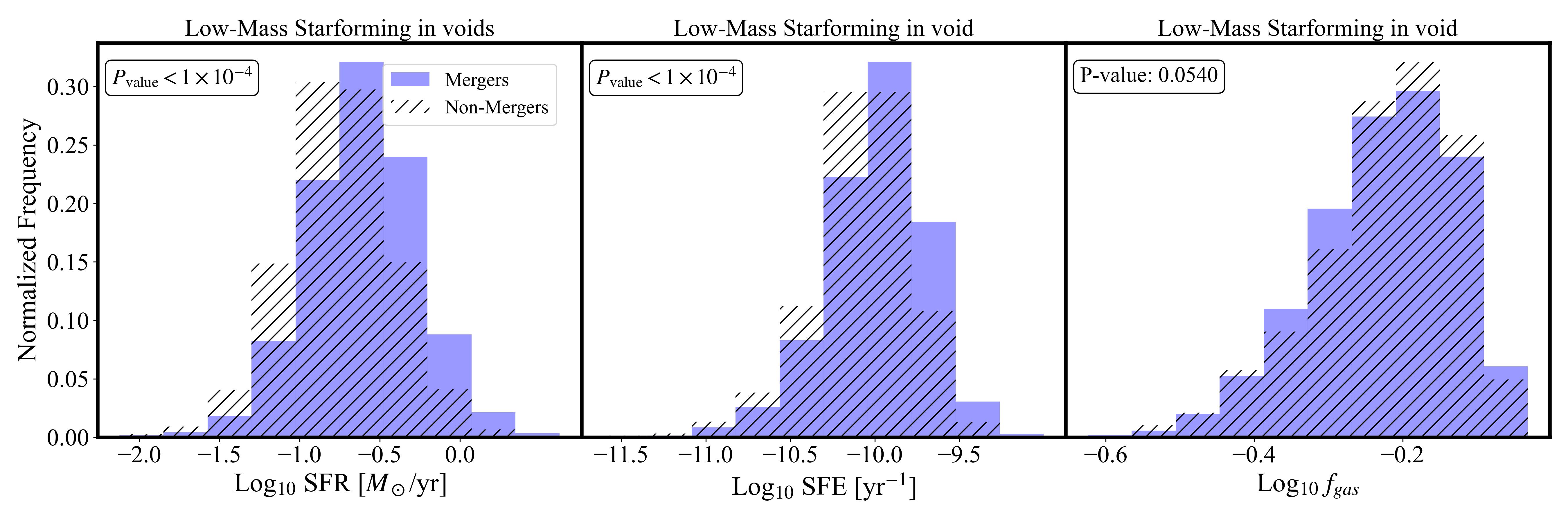}
    
    \vspace{0.1em} 
    
    \includegraphics[width=0.9\textwidth, keepaspectratio]{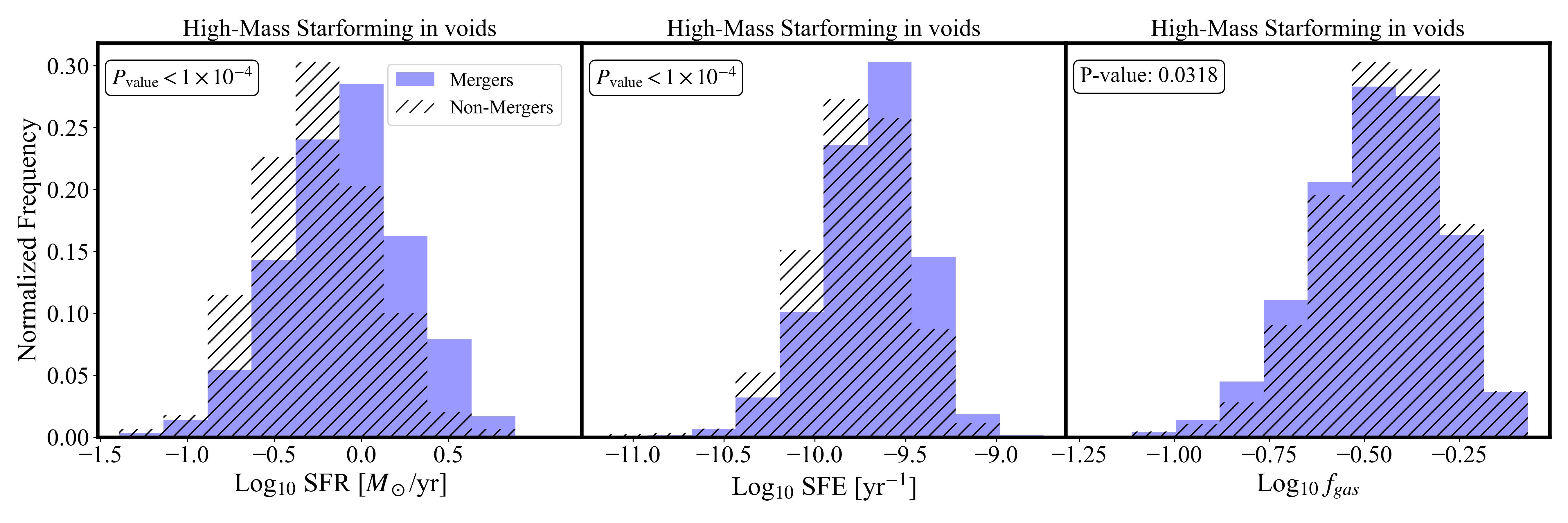}
    
    \caption{Histograms comparing the distributions of star formation rate (SFR), star formation efficiency (SFE), and gas fraction ($f_\text{gas}$) within the last 8 Gyr ($z < 1$) for galaxies in void environments. Systems that experienced at least one merger (Mergers~$\geq 1$) are shown in red and blue, while non-mergers are represented by black hatched histograms. Results are presented from our mass-matched sample for two stellar-mass ranges: low-mass galaxies ($10^{8.5}$--$10^{9.5}\ M_\odot$) and high-mass galaxies ($10^{9.5}$--$10^{10.5}\ M_\odot$). For each stellar-mass bin, we also compute the Kolmogorov–Smirnov (K–S) test to quantify differences between the merger and non-merger populations, reporting the corresponding p-values.
}
    \label{fig:01stellar_mass_comparison020}
\end{figure*}

\begin{figure*}[!t] 
    \centering
    
    \includegraphics[width=0.9\textwidth, keepaspectratio]{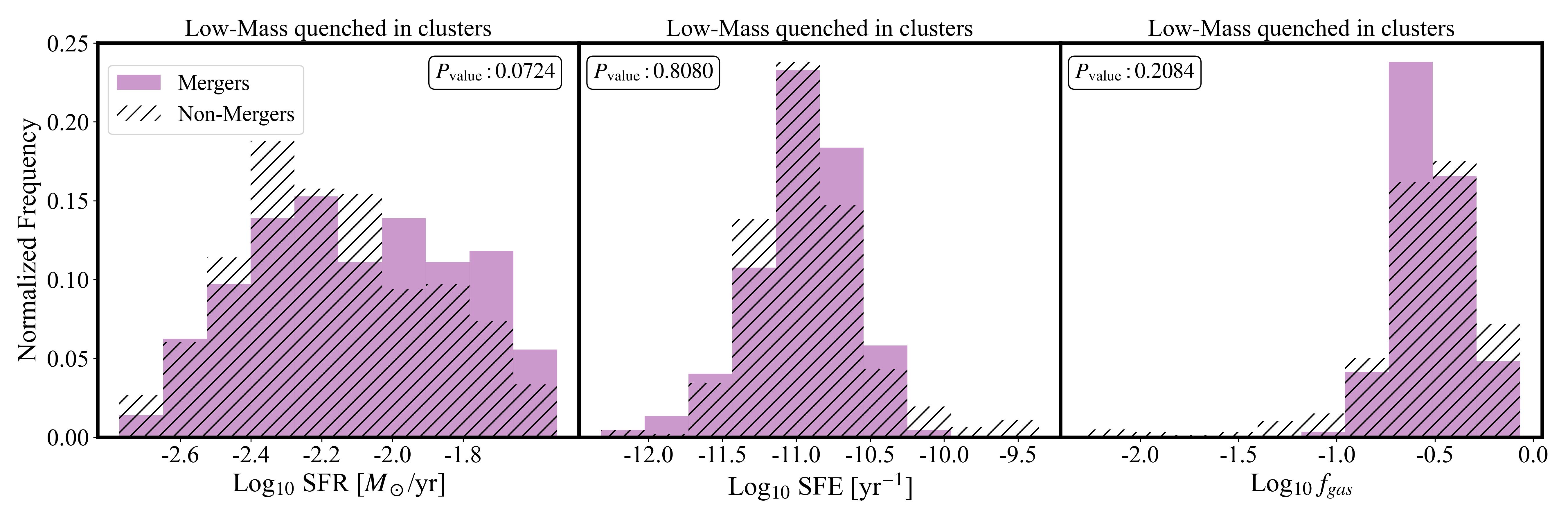}
    
    \vspace{0.1em} 
    
    \includegraphics[width=0.9\textwidth, keepaspectratio]{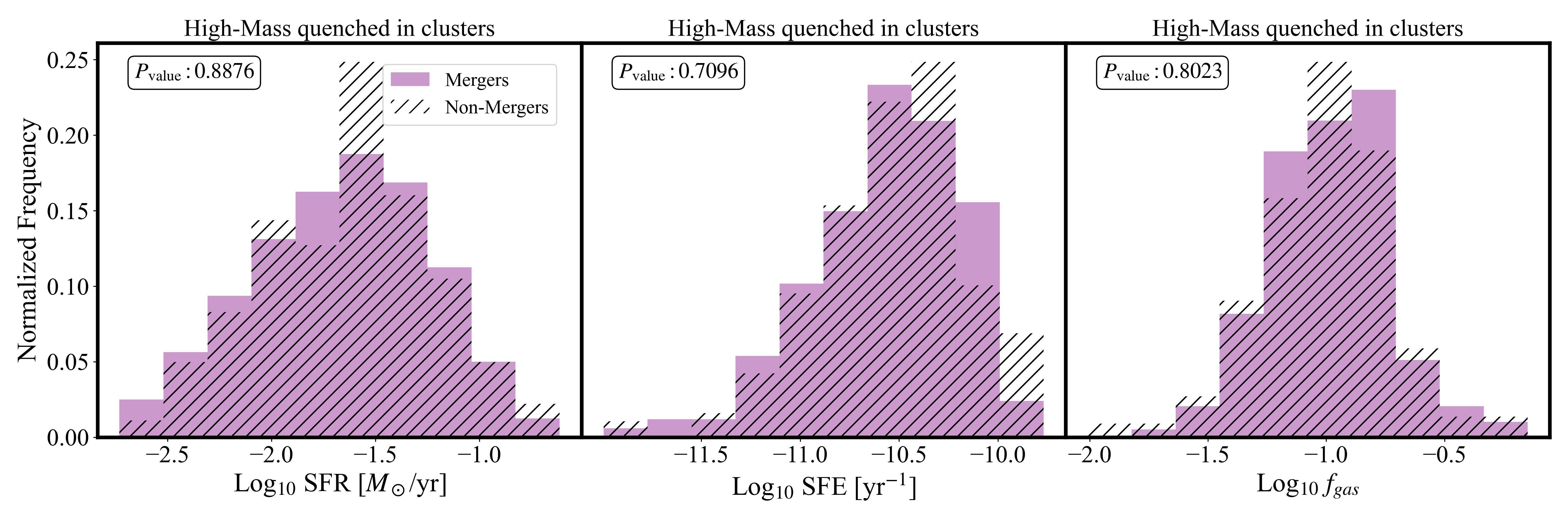}
    
    \vspace{0.1em} 
    
    \includegraphics[width=0.9\textwidth, keepaspectratio]{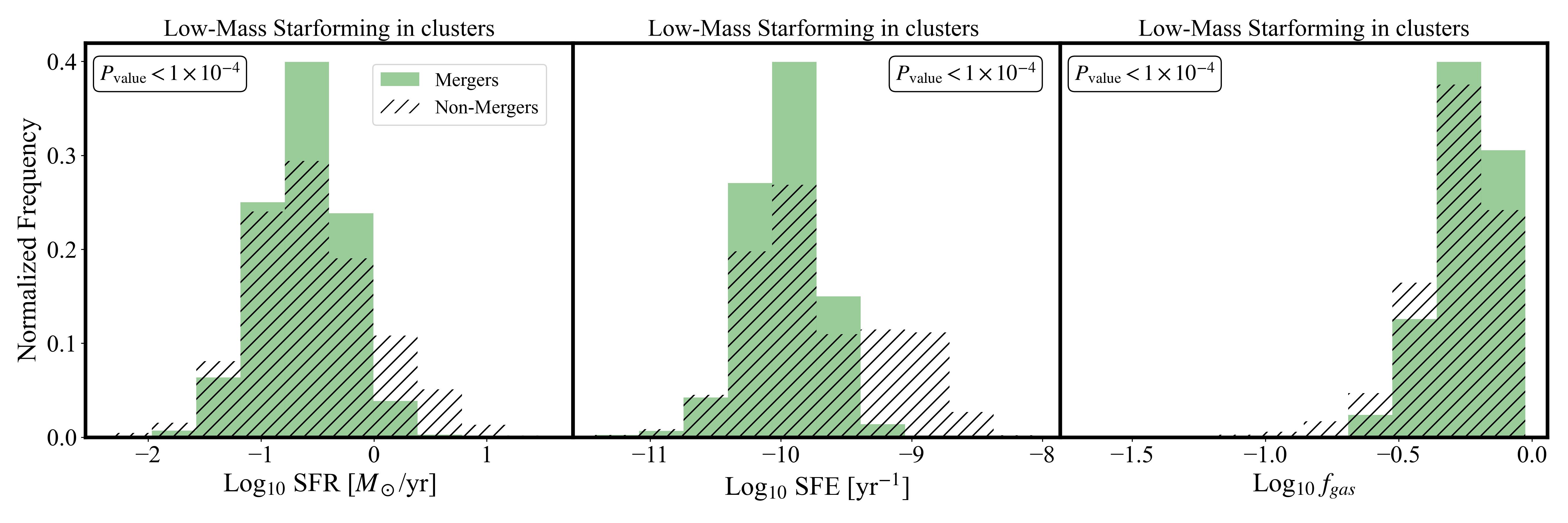}
    
    \vspace{0.1em} 
    
    \includegraphics[width=0.9\textwidth, keepaspectratio]{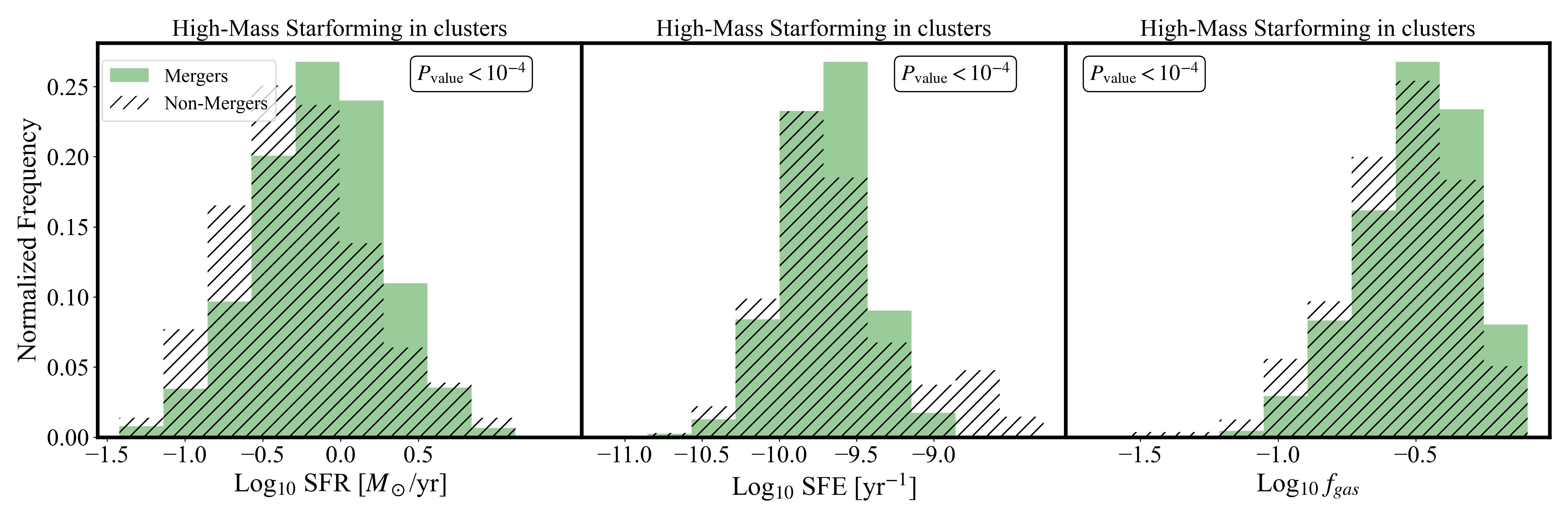}
    
    \caption{Histograms comparing the distributions of star formation rate (SFR), star formation efficiency (SFE), and gas fraction ($f_\text{gas}$) within the last 8 Gyr ($z < 1$) for galaxies in cluster environments. Systems that experienced at least one merger (Mergers~$\geq 1$) are shown in green and purple, while non-mergers are represented by black hatched histograms. Results are presented from our mass-matched sample for two stellar-mass ranges: low-mass galaxies ($10^{8.5}$--$10^{9.5}\ M_\odot$) and high-mass galaxies ($10^{9.5}$--$10^{10.5}\ M_\odot$). For each stellar-mass bin, we also compute the Kolmogorov–Smirnov (K–S) test to quantify differences between the merger and non-merger populations, reporting the corresponding p-values.
}
    \label{fig:01stellar_mass_comparison030}
\end{figure*}

Figure~\ref{fig:01stellar_mass_comparison010} compares the redshift of the last major and minor mergers for star-forming and quenched galaxies, directly contrasting void and cluster environments within each galaxy type. The box plots show the redshift distributions across stellar mass bins, with the horizontal white lines marking the median redshift of the last merger, while the bar charts below each panel show the absolute difference in redshift (ADR) between void and cluster environments for the corresponding galaxy type. Overall, quenched galaxies experience their last mergers systematically earlier (i.e., at higher redshift) than star-forming galaxies, in both voids and clusters: for example, in voids, quenched galaxies have last-major-merger medians of $z \sim 2.2$--$3.3$ across mass bins, compared to $z \sim 1.1$--$1.9$ for star-forming galaxies in the same environment; in clusters, quenched galaxies show medians of $z \sim 1.8$--$3.3$, compared to $z \sim 1.3$--$2.0$ for star-forming galaxies. This is consistent with the earlier mass assembly of quenched galaxies discussed in relation to Figure~\ref{fig:stellar_mass_6comparison}, and holds in both environments.

Turning to the environmental comparison shown directly in Figure~\ref{fig:01stellar_mass_comparison010}, star-forming galaxies show only a small and roughly mass-independent difference between voids and clusters, with ADR $\sim 0.0$--$0.2$ for major mergers and $\sim 0.15$--$0.17$ for minor mergers, indicating that environment has little impact on the timing of the last merger for this population. In contrast, quenched galaxies show a substantially larger environmental effect, with ADR values of $\sim 0.6$--$1.5$ for major mergers and $\sim 0.4$--$0.8$ for minor mergers. For major mergers, quenched galaxies in voids have systematically earlier (higher-redshift) last mergers than their cluster counterparts in the three lower stellar-mass bins ($8.5 \leq \log M_\star/M_\odot < 10.0$), with the largest offset in the lowest-mass bin ($z \sim 3.3$ in voids versus $z \sim 1.8$ in clusters). This trend reverses in the highest stellar-mass bin ($10.0 \leq \log M_\star/M_\odot \leq 10.5$), where quenched cluster galaxies instead show the earlier last major merger ($z \sim 3.3$, compared to $z \sim 2.2$ in voids), suggesting that at the highest masses considered here, cluster environments may drive earlier merger activity in quenched systems, possibly connected to more advanced group/cluster assembly at these masses. A similar reversal at the highest stellar-mass bin is seen for minor mergers, where quenched cluster galaxies again show a slightly earlier last merger ($z \sim 1.9$) than their void counterparts ($z \sim 1.5$), in contrast to the trend at lower masses. This indicates that quenched galaxies in voids and clusters follow more distinct and mass-dependent merger histories than their star-forming counterparts, for which the environment plays a comparatively minor and consistent role in setting the timing of the last merger.

For completeness, a complementary comparison of star-forming versus quenched galaxies within each environment separately is provided in Appendix~\ref{appendix:last_merger_sf_q} (Fig.~\ref{fig:last_merger_appendix}). A detailed comparison of merger fractions across stellar-mass bins and environments is presented in Appendix~\ref{appendix:Merger fractions across stellar-mass bins} (Figure~\ref{fig:01stellar_mass_comparison011}). In summary, star-forming galaxies in voids show the highest major-merger fractions across all mass ranges. Star-forming systems generally undergo more mergers than quenched ones, with this difference being most pronounced in voids. Minor mergers follow a similar pattern, while mini mergers occur more frequently in void galaxies and in higher stellar-mass bins, indicating a strong environmental dependence on merger activity.
\begin{figure*}[!t]
    \centering
    \includegraphics[width=\textwidth, keepaspectratio]{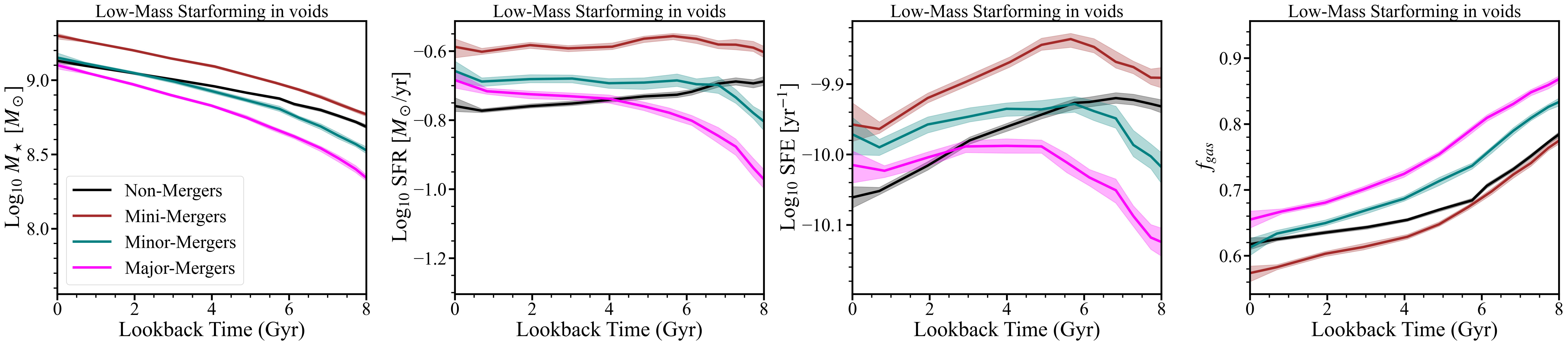}
    \vspace{2.5em}
    \includegraphics[width=\textwidth, keepaspectratio]{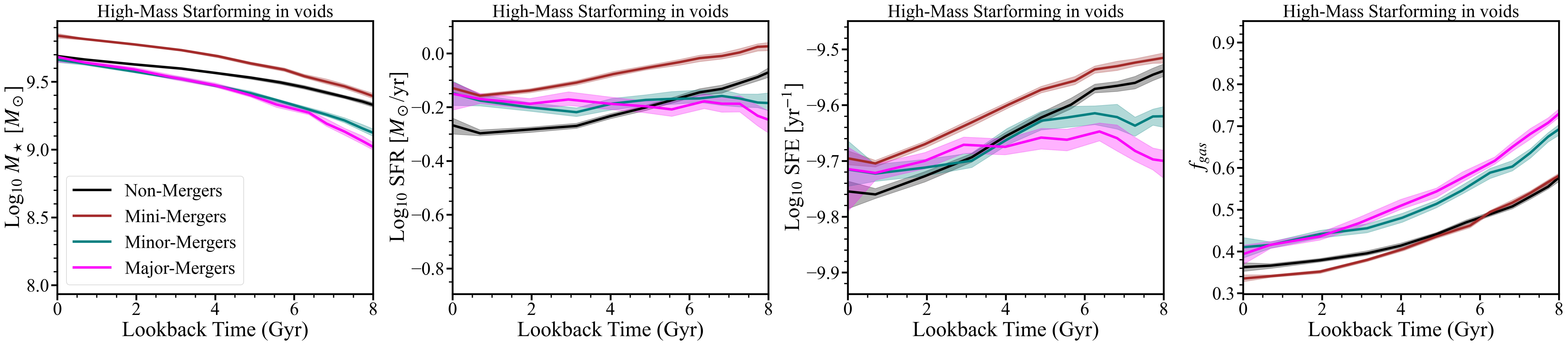}
    \caption{Evolution of stellar mass ($M_\star$; left panels), star formation rate (SFR; second panels), star formation efficiency ($\mathrm{SFE} = \mathrm{SFR}/M_{\rm gas}$; third panels), and gas fraction ($f_{\rm gas} = M_{\rm gas}/(M_\star + M_{\rm gas})$; right panels) as a function of lookback time over the last $\sim 8$ Gyr, for star-forming galaxies in voids. Top: low-mass galaxies ($10^{8.5}$--$10^{9.5}\,M_\odot$); bottom: high-mass galaxies ($10^{9.5}$--$10^{10.5}\,M_\odot$). Galaxies are separated by merger history into non-mergers (black), only mini mergers (brown), only minor mergers (teal), and only major mergers (magenta). Solid lines represent median values; shaded regions show $95\%$ bootstrap confidence intervals.}
\label{fig:01stellar_mass_comparison031}
\end{figure*}

\begin{figure*}[!t]
    \centering
    \includegraphics[width=\textwidth, keepaspectratio]{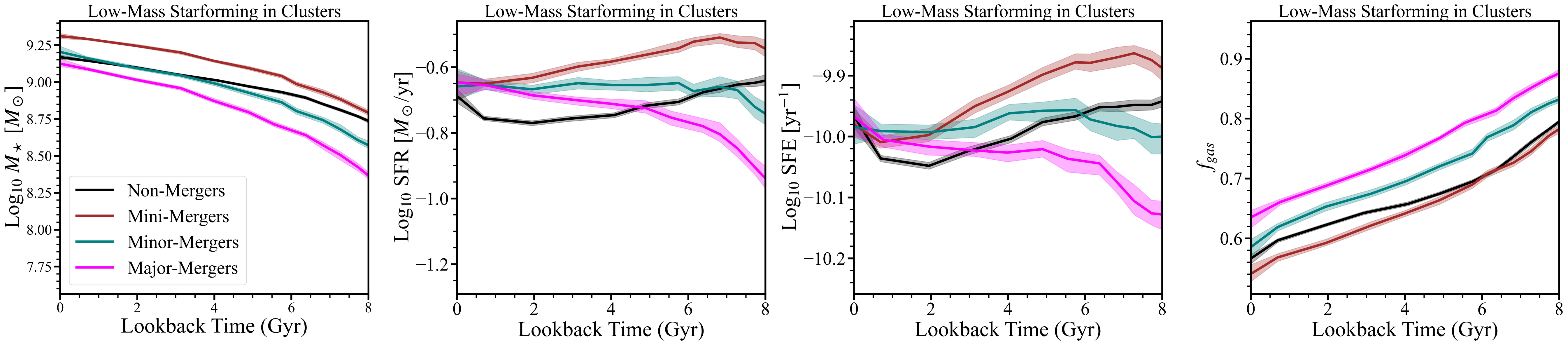}
    \vspace{2.5em}
    \includegraphics[width=\textwidth, keepaspectratio]{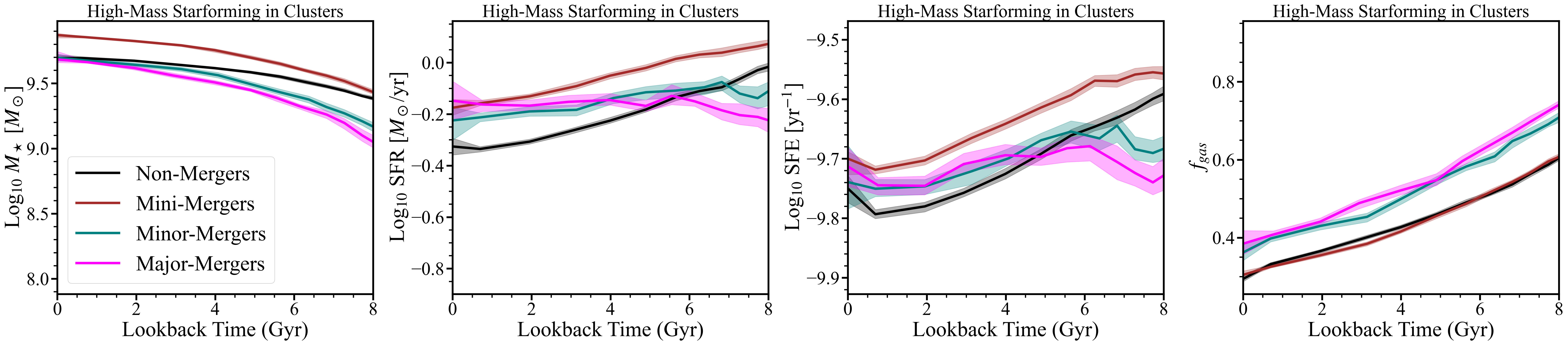}
 \caption{Same as Fig.~\ref{fig:01stellar_mass_comparison031}, but for star-forming galaxies in clusters. Top: low-mass galaxies ($10^{8.5}$--$10^{9.5}\,M_\odot$); bottom: high-mass galaxies ($10^{9.5}$--$10^{10.5}\,M_\odot$). As in Fig.~\ref{fig:01stellar_mass_comparison031}, panels from left to right show stellar mass ($M_\star$), star formation rate (SFR), star formation efficiency (SFE), and gas fraction ($f_{\rm gas}$) as a function of lookback time over the last $\sim 8$ Gyr, with galaxies separated by merger history into non-mergers (black), only mini mergers (brown), only minor mergers (teal), and only major mergers (magenta). Solid lines show median values; shaded regions show $95\%$ bootstrap confidence intervals.}
\label{fig:01stellar_mass_comparison031b}
\end{figure*}

ones, with this difference being most pronounced in voids. Minor mergers follow a similar pattern, while mini mergers occur more frequently in void galaxies and in higher stellar-mass bins, indicating a strong environmental dependence on merger activity.

\subsection{Mean stellar mass of mergers}
An important aspect of our merger analysis is the characteristic stellar mass scale of accreted systems. To quantify this, we consider the MeanStellarMass (last 8 Gyr) from the merger catalogues of \citet{rodriguez2017role} and \citet{eisert2023ergo}. This quantity is defined as the stellar-mass-weighted average of secondary progenitors, where each merger event is weighted by the maximum stellar mass of the infalling galaxy. It therefore captures the typical mass scale of merger companions contributing to galaxy growth over the last 8 Gyr. Figure~\ref{fig:mean_stellar_mass_ratio} presents the normalized distributions of this quantity for star-forming and quenched galaxies in void and cluster environments for our mass-matched samples. The mean values are $\log_{10}(M/M_\odot) \approx 7.9$ in voids and $\approx 8.0$ in clusters for star-forming galaxies, compared to $\approx 7.75$ and $\approx 7.95$ for quenched galaxies. In void environments, star-forming galaxies exhibit systematically higher mean merger masses than quenched systems, with an offset of $\approx 0.15$ dex. 

In contrast, the difference in clusters is small ($\approx 0.05$ dex), and the two populations show nearly identical mean values. This indicates that the star-forming--quenched distinction in merger mass scale is substantially stronger in low-density environments. Comparing environments within each galaxy type reveals a clear difference. Star-forming galaxies show nearly identical distributions in voids and clusters, with only a small shift of $\approx 0.10$ dex, indicating little environmental dependence. In contrast, quenched galaxies show a larger environmental shift, with mean values increasing from voids to clusters by $\approx 0.20$ dex. This suggests that the merger histories of quenched galaxies may be more sensitive to environment than those of star-forming galaxies, though we note below that this difference does not reach formal statistical significance given our sample sizes.

To assess the statistical significance of these differences, we performed pairwise Kolmogorov--Smirnov (K--S) tests. The difference between star-forming galaxies in voids and clusters is small but statistically significant ($D=0.054$, $p \approx 10^{-6}$). The difference between quenched galaxies in voids and clusters, despite being comparable in magnitude to the star-forming case, does not reach significance at the 0.05 level ($D=0.153$, $p=0.065$), likely reflecting the modest sample size of quenched void galaxies ($N=91$, compared to $N=325$ in clusters). Comparisons between star-forming and quenched galaxies within a single environment are only marginal in voids ($p=0.054$) and not significant in clusters ($p=0.456$). We therefore caution that the apparent environmental dependence of merger mass scale for quenched galaxies, while suggestive, should be interpreted with care given the limited sample size.

We tested the statistical robustness of an apparent bimodal structure in the mean merger mass distribution of quenched galaxies using Hartigan's dip test and a Gaussian Mixture Model (GMM) comparison (1 vs. 2 components, compared via the Bayesian Information Criterion). The dip test finds no significant evidence against unimodality for either population (voids: $p=0.84$; clusters: $p=0.99$), and while a 2-component GMM is formally preferred over a single Gaussian in both cases, the improvement is only marginal ($\Delta\mathrm{BIC} = 2.3$ in voids and $5.0$ in clusters), below the threshold typically considered strong evidence, and of comparable magnitude in both environments. We therefore do not find robust statistical support for a bimodal distribution specific to quenched galaxies in voids, and do not interpret this feature further.

\subsection{Mergers vs. Non-Mergers: Galaxy Properties}

To investigate the impact of mergers on star formation and quenching galaxies, we conducted a targeted study comparing two distinct groups: galaxies with no mergers (Non-mergers) and galaxies with at least one merger (Mergers) ($N_{\mathrm{mergers}} \geq 1$). In this study, mergers are broadly defined to include any mergers across any stellar mass ratio within the last 8 Gyr ($z < 1$), a period during which environmental effects play a significant role. In Figures~\ref{fig:01stellar_mass_comparison020} and \ref{fig:01stellar_mass_comparison030}, we compare the distributions of star formation rates (SFRs), star formation efficiency ($\mathrm{SFE} = \mathrm{SFR} / M_{\mathrm{gas}}$), and gas fraction ($f_{\mathrm{gas}} = M_{\mathrm{gas}} / (M_{\mathrm{gas}} + M_\star)$). Galaxies with at least one merger (hereafter mergers) are shown in red, blue, purple, and green, while non-merger galaxies are shown as black hatched histograms. The distributions are presented separately for star-forming and quenched galaxies in void and cluster environments, respectively. 

To ensure a detailed comparison, the galaxy samples are divided into two mass bins: low-mass ($10^{8.5}$--$10^{9.5}\,M_\odot$) and high-mass ($10^{9.5}$--$10^{10.5}\,M_\odot$). The final sample sizes for each merger/non-merger subgroup, environment, and mass bin are reported in Table~\ref{tab:merger_nonmerger_samples}. We also performed the Kolmogorov--Smirnov (K--S) test, calculating the K--S statistic and the p-value on each low- and high-mass population of star-forming and quenched galaxies in both environments to determine the significance of the differences. We note that the quenched-void samples are considerably smaller than the other subgroups (N=86 in the low-mass bin, N=67 in the high-mass bin; see Table~\ref{tab:merger_nonmerger_samples}), which limits the statistical power of the K--S test for this population; the reported non-significant differences for quenched void galaxies discussed below should therefore be interpreted with some caution, as they may partly reflect limited sample size rather than a genuine absence of merger impact.

 In Figure~\ref{fig:01stellar_mass_comparison020}, the first panel compares low-mass quenched galaxies in voids that experienced at least one merger with those that had no mergers within the last 8 Gyr ($z < 1$). The distributions of star formation rate (SFR), star formation efficiency (SFE), and gas fraction show no statistically significant differences, with high p-values indicating that mergers do not appreciably influence the properties of low-mass quenched void galaxies. High-mass quenched galaxies in voids display a similar pattern: most properties show substantial overlap between mergers and non-mergers, suggesting only a limited merger impact. The SFR distribution, for example, yields a p-value of 0.50. A moderate difference emerges only in SFE, where the K–S test gives $p < 0.05$, implying that mergers may introduce a modest change in how efficiently gas is converted into stars, although the effect remains secondary. In contrast, star-forming galaxies in voids show clear merger-driven differences. For low-mass systems, mergers substantially elevate both SFR and SFE ($p < 10^{-4}$), indicating enhanced star formation activity and a more efficient conversion of gas into stars. The gas fraction, however, shows only a marginal difference ($p = 0.05$), suggesting a weaker merger influence on the overall gas reservoir. High-mass star-forming galaxies follow the same trend: mergers strongly affect SFR and SFE ($p < 10^{-4}$), while the gas fraction exhibits only mild variation ($p = 0.03$), implying that the primary impact of mergers in void environments is on star formation activity and efficiency rather than on gas content.

In Figure~\ref{fig:01stellar_mass_comparison030}, low-mass quenched galaxies in clusters show minimal differences between mergers and non-mergers across the examined properties, indicating that mergers have a limited impact on this population. The SFR distribution yields a p-value of 0.07, suggesting only a weak difference between the two groups. For high-mass quenched galaxies, the K–S test similarly shows no significant merger effects: the SFE distribution displays no meaningful variation ($p = 0.70$), and the gas fraction also shows negligible differences, confirming that mergers exert little influence on the physical properties of quenched systems. In contrast, star-forming galaxies in clusters exhibit strong merger-driven differences. For low-mass systems, mergers significantly enhance both SFR and SFE ($p < 10^{-4}$). Unlike the trends observed in void environments, the gas fraction in clusters is also strongly affected by mergers ($p < 10^{-4}$). High-mass star-forming galaxies follow the same pattern: mergers substantially modify SFR and SFE ($p < 10^{-4}$), while also producing pronounced differences in gas content ($f_{\mathrm{gas}}$: $p < 10^{-4}$).

Interestingly, our analysis reveals no significant differences between mergers and non-mergers in the properties of quenched galaxies in both environments suggesting that once star formation has ceased, mergers do not substantially modify their star-forming properties. However, while mergers exert a notable impact on the properties of star-forming galaxies in both voids and clusters, their effect on gas content reveals a significant contrast: mergers have an almost negligible influence on the gas fraction in star-forming galaxies in voids, whereas they significantly affect the gas fraction in star-forming galaxies in cluster environments. As a result, the impact of mergers on gas content can vary significantly depending on the surrounding environment, and mergers can yield different results in different environments for the gas content of star-forming galaxies. The differences between quenched and star-forming galaxies in the effects of mergers on their main properties possibly emphasize the role of the galactic gas content. The gas fraction of galaxies may be effective during mergers. The gas fraction is not the sole driver, albeit a strong driver of SFR enhancement in galaxy mergers. In addition, the possible effects of feedback efficiency and galaxy morphology could contribute to the discrepancy between star-forming and quenched galaxies \citep{hani2020interacting}. For instance, mergers of bulge-dominated galaxies have been predicted \citep[e.g.,][]{cox2008effect} and observed \citep[e.g.,][]{saintonge2012impact} to exhibit lower star formation rate (SFR) enhancements. However, exploring the interplay between galaxy morphology and SFR enhancements in mergers within IllustrisTNG is beyond the scope of this work. 

\subsection{Effects of Merger Types on Star-forming Galaxies}

For a more detailed analysis of the effects of mergers on star-forming galaxies in different environments, we consider in Fig.~\ref{fig:01stellar_mass_comparison031} and Fig.~\ref{fig:01stellar_mass_comparison031b}  galaxies selected from our mass-matched sample in two stellar mass ranges: low-mass ($10^{8.5}$--$10^{9.5}\,M_\odot$) and high-mass ($10^{9.5}$--$10^{10.5}\,M_\odot$). Within each mass bin, galaxies are further classified according to their merger history over the last $\sim 8$ Gyr into systems experiencing exclusively mini, minor, or major mergers, as well as non-merger galaxies. We note that these categories are not mutually exclusive by construction but rather reflect the natural variance in merger histories: since mini mergers (mass ratio $<1/10$) are intrinsically far more frequent than minor or major mergers, a consequence of the steep subhalo mass function \citep{rodriguez2015merger}, many galaxies accumulate several mini mergers without experiencing any minor or major merger within the same 8 Gyr window, and vice versa. Similarly, ``non-merger'' galaxies in this window may have experienced mergers earlier in their history (at lookback times $>8$ Gyr) but had a quiescent accretion history over the period considered here. We then examine how these different merger channels affect key galaxy properties, including the star formation rates (SFRs), the star formation efficiency ($\mathrm{SFE} = \mathrm{SFR}/M_{\rm gas}$), and the gas fraction ($f_{\rm gas} = M_{\rm gas}/(M_\star + M_{\rm gas})$). Quenched galaxies are excluded from this analysis, as their star-forming properties (e.g., SFR and SFE) show little or no dependence on merger activity. We therefore restrict our study to star-forming systems, for which mergers have a measurable impact on star formation and gas-related properties.

In Fig.~\ref{fig:01stellar_mass_comparison031} and Fig.~\ref{fig:01stellar_mass_comparison031b}, the solid lines represent the median values of each quantity in bins of lookback time, while the shaded regions indicate the associated uncertainties. These uncertainties are estimated using bootstrap resampling, with the shaded bands corresponding to the $95\%$ confidence intervals derived from the 2.5th and 97.5th percentiles of the bootstrap distributions.

Before examining the SFR, SFE, and gas-fraction trends, we first inspect the evolution of stellar mass for each merger-category subsample, shown in the leftmost panel of each figures. In both voids and clusters, and for both mass bins, stellar mass grows only mildly over the 8~Gyr interval considered (by $\sim 0.4$--$0.7$ dex), reflecting the fact that our mass-matched samples are already restricted to a narrow stellar-mass range at $z=0$ and are selected to remain star-forming throughout. Notably, mini-merger galaxies consistently occupy the higher end of the stellar-mass distribution at all lookback times relative to non-mergers, minor-, and major-merger galaxies, in both environments and mass bins. This context is important for interpreting the SFR trends below: because the stellar mass of mini-merger galaxies changes only modestly over this period, an approximately constant SFR does not imply an anomalously flat specific star formation rate, but rather reflects the limited dynamic range in stellar mass growth within our mass-matched, star-forming-selected samples.

For low-mass star-forming galaxies in voids, the influence of different merger channels evolves with lookback time. In the SFR panel, mini mergers dominate the early evolution, maintaining the highest SFR across most of the lookback time interval. As shown in the stellar mass panel of both figures, this near-constant SFR is consistent with the correspondingly modest growth in stellar mass for this subsample over the same interval, rather than indicating an anomalous evolutionary behaviour. At earlier epochs, however, major and minor mergers show an opposite trend relative to mini mergers, tending to suppress the SFR compared to the non-merging population. This suppression is stronger for major mergers than for minor mergers. Toward more recent epochs, the influence of minor and major mergers becomes more apparent. Minor mergers appear to affect the SFR over a longer interval, roughly up to $t_{\rm lookback}\lesssim6$~Gyr, whereas the impact of major mergers becomes more pronounced at later times, particularly for $t_{\rm lookback}\lesssim4$~Gyr. A similar behavior is observed for the SFE. Mini mergers are particularly effective at earlier epochs, reaching a peak around $t_{\rm lookback}\sim6$~Gyr. At more recent times, both minor and major mergers contribute more noticeably to the evolution of SFE, although their enhancement remains weaker than the earlier increase associated with mini mergers. In contrast, the gas fraction shows a different pattern. Mini mergers appear to have little impact on $f_{\rm gas}$, with values comparable to or slightly lower than those of other merger channels. The highest gas fractions are generally associated with major mergers, while minor mergers show a more moderate effect. These trends suggest that major mergers play a more significant role in regulating the gas reservoir in low-mass void galaxies. Moreover, their influence appears to vary with cosmic time, showing a suppressing effect on star formation at earlier epochs but a more positive contribution at recent times, similar to the behavior seen for minor mergers.

In comparison to low-mass galaxies in clusters, we find broadly similar trends in the influence of minor and major mergers: both channels show a modest enhancement of SFR and SFE toward recent epochs, while having a suppressing or neutral effect at earlier times. However, the behavior of mini mergers differs between environments. In clusters, the impact of mini mergers decreases toward the present time, whereas in void galaxies their effect remains approximately constant across the lookback time interval. An additional feature in cluster galaxies is a noticeable early peak in SFE at $t_{\rm lookback}\sim7.5$~Gyr, occurring earlier than in the void population. Moreover, non-merging galaxies in clusters exhibit an increase in both SFR and SFE at late times ($t_{\rm lookback}\lesssim2$~Gyr), which may reflect the influence of the cluster environment on the gas supply and the regulation of star-formation activity.

For high-mass star-forming galaxies in both voids and clusters, we observe broadly similar trends. Mini mergers appear to be more effective in enhancing SFR and SFE than other merger types. In contrast, major and minor mergers show only a weak influence on SFR and SFE at earlier times ($t_{\rm lookback}\sim6$--$8$~Gyr). Overall, these effects are weaker than those observed in low-mass systems. However, at later times ($t_{\rm lookback}\lesssim6$~Gyr), mergers become more efficient in enhancing star-formation activity. In terms of gas content, both minor and major mergers exhibit a noticeable impact, with their effects on the gas fraction being approximately comparable.

We find that the influence of mergers depends on both stellar mass and environment. For low-mass galaxies, merger-driven effects are more pronounced in voids, where a clear separation between different merger channels is maintained. In clusters, this contrast is reduced, largely due to the late-time rise of the non-merging population. In contrast, high-mass galaxies exhibit broadly similar trends in both environments, with weaker distinctions between merger classes. This indicates that environmental modulation of merger-driven evolution is most significant for low-mass systems, while high-mass galaxies are largely insensitive to environmental differences. Across all masses and environments, mini mergers are the most effective channel for enhancing SFR and SFE relative to non-mergers.

Minor and major mergers show a moderate enhancement that becomes increasingly important toward recent epochs; both channels are comparatively weak at early times but can approach or even exceed the non-merging population at late times. The behaviour of the gas fraction differs from that of SFR and SFE: in low-mass galaxies, major mergers dominate the gas reservoir, whereas in high-mass systems both minor and major mergers contribute substantially. We note that, in contrast to the pronounced environmental dependence found for quenched galaxies in Sects.~3.4 and 3.5, the merger-driven evolution of star-forming galaxies is comparatively similar in voids and clusters, particularly at high stellar mass. This similarity is itself an informative result: it indicates that, for star-forming systems, the internal response to a given merger event is governed primarily by the merger's mass ratio and cosmic epoch rather than by the large-scale environment, complementing the strong void--cluster contrast identified for quenched galaxies elsewhere in this work.

\section{Conclusions }

This study utilizes the high-resolution TNG300\_1 simulation from the IllustrisTNG project \citep{springel2018first, pillepich2018first} to analyze galaxies with stellar masses in the range \(10^{8.5}\,M_{\odot} \leq M_{\star} \leq 10^{10.5}\,M_{\odot}\) to compare the evolutionary histories of star-forming and quenched galaxies in dense (groups \& clusters, \(10^{13} \leq M_{200} < 10^{15} \, M_\odot\)) and under-dense (voids) environments and investigate the evolutionary path of the main properties of these galaxies over the past  \( \approx 10.5Gyr\) (\(z < 2\)). Additionally, we investigated the mass assembly histories of star-forming and quiescent galaxies in these environments. This research represents the first statistical examination of mergers—distinguishing between major, minor, and mini mergers—in different environments across various evolutionary stages of star-forming and quenched low-mass galaxies, motivated by the observational challenges in investigating mergers and evolutionary trajectories in these low-mass systems. Our main results are as follows: 

(1) At $z=0$, star-forming galaxies in voids and clusters show broadly similar properties, with only minor environmental differences. In contrast, quenched galaxies exhibit much stronger environmental dependence, especially in gas content, dark-matter mass, and gas number density, indicating fundamentally different quenching processes in voids and clusters in the same stellar mass. Interestingly, quenched void galaxies contain higher dark-matter masses across all stellar mass bins compared to other samples. 
 
(2) Our analysis of the evolutionary pathways indicates that star-forming galaxies evolve largely independently of their environment. Quenched galaxies, however, while showing some similar evolutionary trends such as decreasing SFR below $z \approx 0.5$ regardless of environment, exhibit significant differences in the evolution of their dark matter mass, gas mass, and gas fraction. These distinctions highlight that the quenching mechanisms in dense and underdense regions are fundamentally different.

(3) Quenched galaxies assemble their stellar mass earlier and more rapidly than star-forming galaxies, independent of environment, likely due to intense early star formation that depleted their gas reservoirs and led to quenching by \(z = 0\) \citep{behroozi2019universemachine}. While star-forming galaxies exhibit similar mass assembly histories in voids and clusters, quenched galaxies diverge significantly, particularly above 50\% of their final mass, highlighting environmental effects at \(z < 1\). In these low redshifts, quenched cluster galaxies form earlier than their void counterparts, especially at higher stellar masses, suggesting that the processes leading to quenching in clusters may also accelerate the mass assembly of galaxies. These findings are consistent with the classification of void star formation histories into short- and long-timescale types \citep{dominguez2023galaxies}, and we propose similar patterns may exist in clusters, albeit with different quenching mechanisms. We emphasize that these results apply to galaxies in the stellar mass range \(10^{8.5} - 10^{10.5} \, M_\odot\), where star formation is still active and quenching pathways are environmentally sensitive. 

(4) Quenched void galaxies are not subject to environmental quenching mechanisms such as strangulation or ram pressure stripping \citep{rodriguez2024evolutionary}. However, they consistently reside in more massive dark matter halos within twice their stellar half-mass radius over the past 8 Gyr, especially at higher stellar masses. This may enhance star formation efficiency and accelerate gas consumption, leading to declining gas content and SFRs over the last 5 Gyr \citep{boylan2025accelerated}.

(5) Our results in analysing merger statistics reveal several key trends (here, $R$ denotes the void-to-cluster merger-fraction ratio introduced in Sect.~3.4.1, with $R>1$ indicating a higher merger fraction in voids and $R<1$ indicating a higher merger fraction in clusters):
\begin{itemize}
\item Merger activity is strongly environment-dependent, with void galaxies experiencing more mergers—especially mini and minor mergers—toward recent times, whereas at earlier epochs the merger histories of void and cluster galaxies are broadly similar.

\item Star-forming galaxies show a smooth and moderate environmental dependence, with merger fractions increasing toward the present day (reaching $R \sim 1.6$--$1.8$) but without the dramatic reversals seen in quenched systems.

\item Quenched galaxies exhibit the strongest environmental contrast, with void quenched galaxies undergoing far more small mergers at late times (mini: $R \sim 3.0$--$3.3$, minor: $R \sim 2.7$--$2.9$), while at intermediate epochs clusters dominate their merger activity ($R \sim 0.6$--$0.8$).

\item Quenched galaxies experience their last mergers systematically earlier than star-forming galaxies in both environments; however, the environmental (void-cluster) dependence of this timing is small and mass-independent for star-forming galaxies, but large and mass-dependent for quenched galaxies, reversing at the highest stellar masses.

\item Mini mergers represent the dominant channel for enhancing SFR and SFE across all masses and environments, whereas major and minor mergers show weaker, more time-dependent effects; major mergers additionally dominate the gas reservoirs of low-mass galaxies.

\end{itemize}

(6) Star-forming galaxies show little environmental dependence in the stellar mass of their merger companions ($\approx 0.10$ dex shift between voids and clusters), a difference that is small but statistically significant. Quenched galaxies show a larger apparent shift toward higher merger masses in clusters ($\approx 0.20$ dex); however, this difference does not reach formal statistical significance given the limited sample size of quenched galaxies in voids ($N=91$), and should be interpreted with caution. We tested for the previously suggested bimodality in the mean merger mass of quenched galaxies using Hartigan's dip test and a Gaussian Mixture Model comparison, and found no robust statistical evidence for a bimodal distribution specific to either environment.

(7) Mergers have little to no measurable impact on quenched galaxies in either voids or clusters, with SFR, SFE, and gas fraction distributions statistically indistinguishable from non-mergers; however, this result should be interpreted with some caution for quenched galaxies in voids, given their limited sample size ($N \sim 90$). In contrast, star-forming galaxies show strong merger-driven enhancements in both SFR and SFE across all environments, though the impact on gas content is environment-dependent: negligible in voids but strongly significant in clusters. These results imply that merger-induced star formation is effective only in gas-rich systems, while quenched (gas-poor or bulge-dominated) galaxies remain largely unaffected.

(8) Merger-driven effects depend on both stellar mass and environment. In low-mass galaxies, these effects are strongest in voids, while in clusters the contrast between merger channels is reduced by the late-time rise of non-mergers. High-mass galaxies show similar trends in both environments, indicating weaker environmental influence; this similarity is itself notable, as it contrasts with the pronounced environmental dependence found for quenched galaxies elsewhere in this work. Mini mergers consistently provide the largest boost to SFR and SFE, whereas minor and major mergers become important mainly at late times. For gas fractions, major mergers dominate in low-mass systems, while both minor and major mergers contribute in high-mass galaxies.

In this work, we find that within the stellar mass ranges considered, star-forming galaxies evolve in broadly similar ways across different environments, whereas quenched galaxies display strong environmental dependencies. All quenched systems tend to form earlier, independent of environment, which may naturally lead to earlier gas consumption. However, the dominant quenching mechanisms differ: in clusters, quenched galaxies are likely shaped by environmental processes such as gas stripping and strangulation, while quenched galaxies in voids typically reside in more massive dark matter haloes and experience fewer recent mergers—particularly major ones—reducing their ability to replenish their gas reservoirs. Mini mergers dominate the merger budget for all galaxy types over the past $\sim 8$~Gyr and are significantly more common in voids, with minor mergers showing a similar but weaker trend. In star-forming galaxies, mergers—especially mini mergers—play an important role in enhancing and sustaining the SFR, whereas major mergers have the strongest impact on increasing the gas fraction, a behaviour most clearly visible in voids. In contrast, mergers have little to no effect on quenched galaxies, likely due to the reduced frequency of late-time mergers in the last $\sim 8$~Gyr. Overall, our results highlight the combined influence of stellar mass, environment, and merger history in driving the divergent evolutionary pathways of galaxies.

\section*{Acknowledgements}
 
We thank the anonymous referee for their insightful and constructive feedback, particularly regarding the comparison of environments, which enhanced the clarity and depth of this study. Their suggestions helped improve our analysis of galaxy evolution and merger histories, leading to a more comprehensive understanding of environmental effects on low-mass galaxies.  
We gratefully acknowledge the IllustrisTNG collaboration for making the TNG300 simulation publicly available. The data used in this work were accessed through the IllustrisTNG database (\url{www.tng-project.org}), and we thank the team for their contributions to developing and maintaining this valuable resource, which underpins the scientific results presented here.

%
%

\bibliographystyle{aa}  
\bibliography{aanda}  

\clearpage              
\appendix               

\section{Additional Merger Statistics}
\label{appendix:Additional Merger Statistics}

\begin{figure*}[t] 
    \centering
    \includegraphics[width=0.95\textwidth, keepaspectratio]{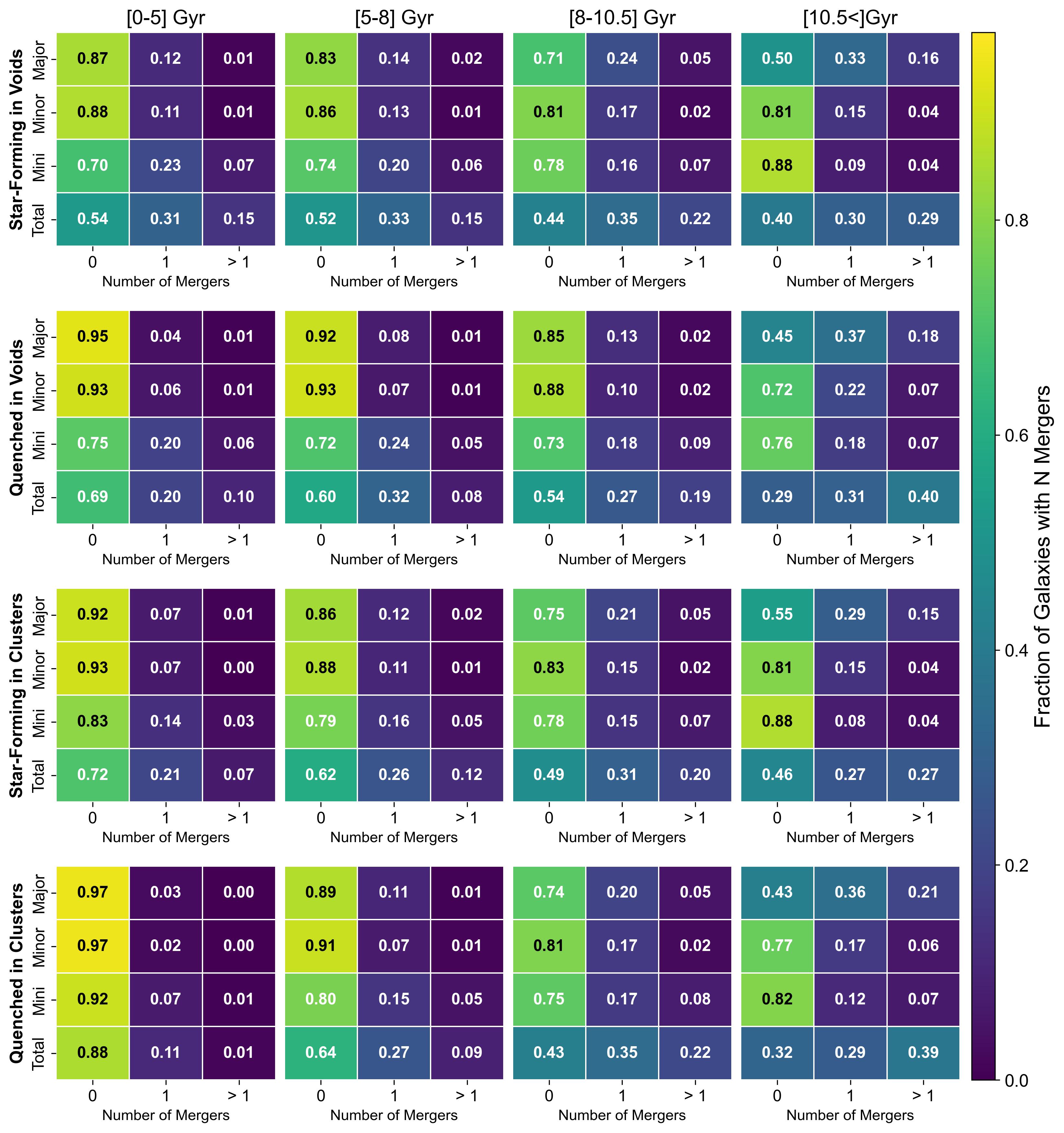}
   \caption{Fraction of galaxies undergoing four types of mergers—major (stellar mass ratio > 1/4), minor(stellar mass ratio between 1/10 and 1/4 ), mini(stellar mass ratio < 1/10), and total (any stellar mass ratio ) mergers—across distinct periods of cosmic evolution: the last 5 Gyr, 5–8 Gyr, 8–10.5 Gyr, and earlier epochs (>10.5 Gyr) for star-forming and quenched galaxy populations in voids and cluster environments. The color gradient illustrates the fraction of galaxies experiencing varying numbers of mergers, categorized as zero mergers, one merger, and multiple mergers (more than one), during these evolutionary stages.
}

    \label{fig:stellar_mass_080comparison}
\end{figure*}

Figure~\ref{fig:stellar_mass_080comparison} summarizes the merger histories of star-forming and quenched galaxies in voids and clusters. At early times (>10.5 Gyr), major mergers dominate for all galaxies, with quenched systems showing slightly higher merger fractions. At later epochs, cluster galaxies of both types follow similar trends, while void galaxies diverge: quenched void galaxies show a sharp decline in major mergers, whereas star-forming void galaxies maintain higher major, minor, and mini merger activity. Mini mergers remain particularly common in voids during the last 8 Gyr. Overall, the heatmap reveals an early epoch with high merger activity across all environments, followed by a later epoch where environmental differences become significant.

\begin{figure*}[hbtp]
    \centering
    
    \includegraphics[width=0.85\textwidth, height=0.5\textheight, keepaspectratio]{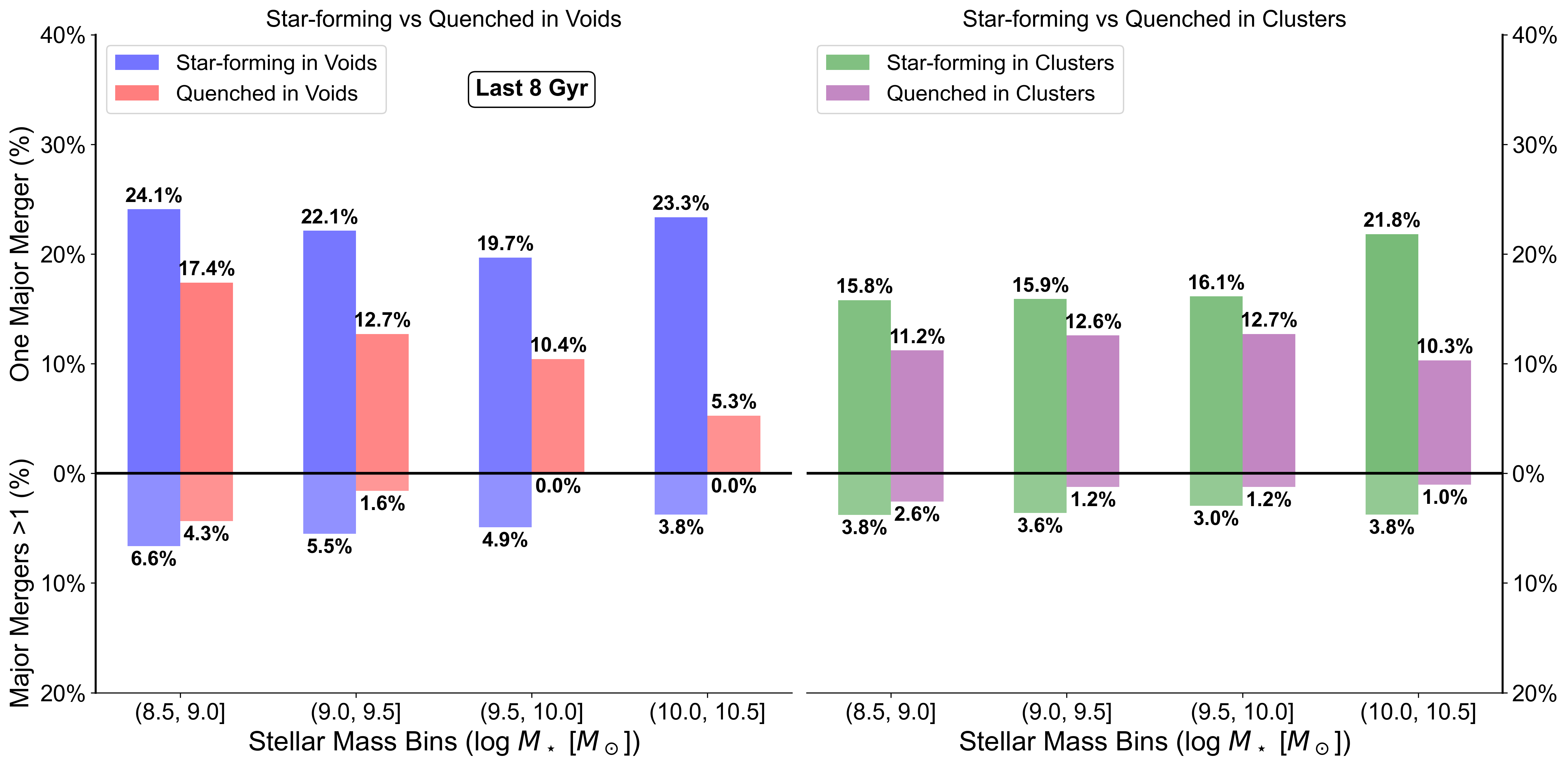}
    
    \vspace{0.2em}
    
    \includegraphics[width=0.85\textwidth, height=0.5\textheight, keepaspectratio]{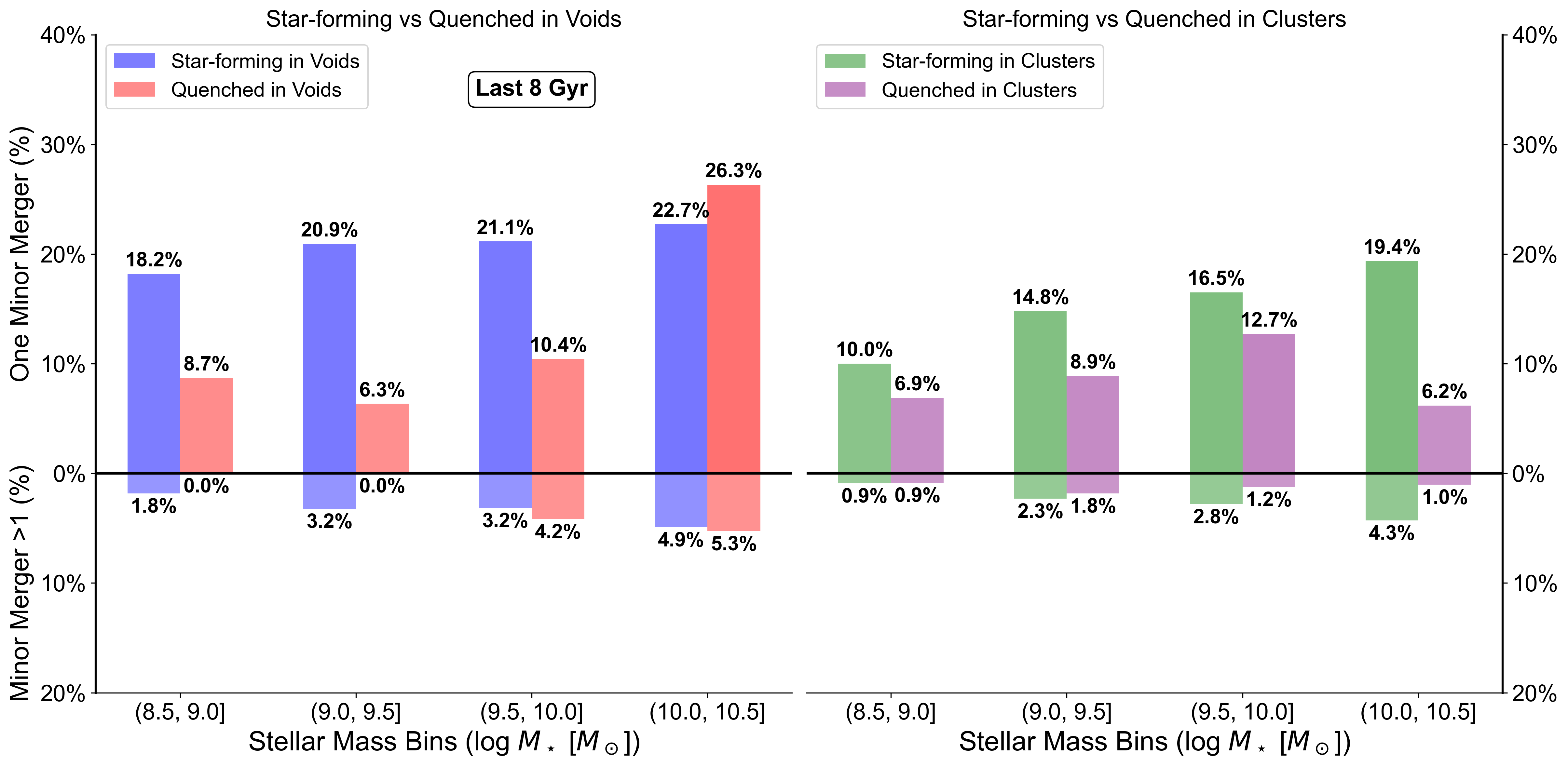}
    
    \vspace{0.2em}
    
    \includegraphics[width=0.85\textwidth, height=0.5\textheight, keepaspectratio]{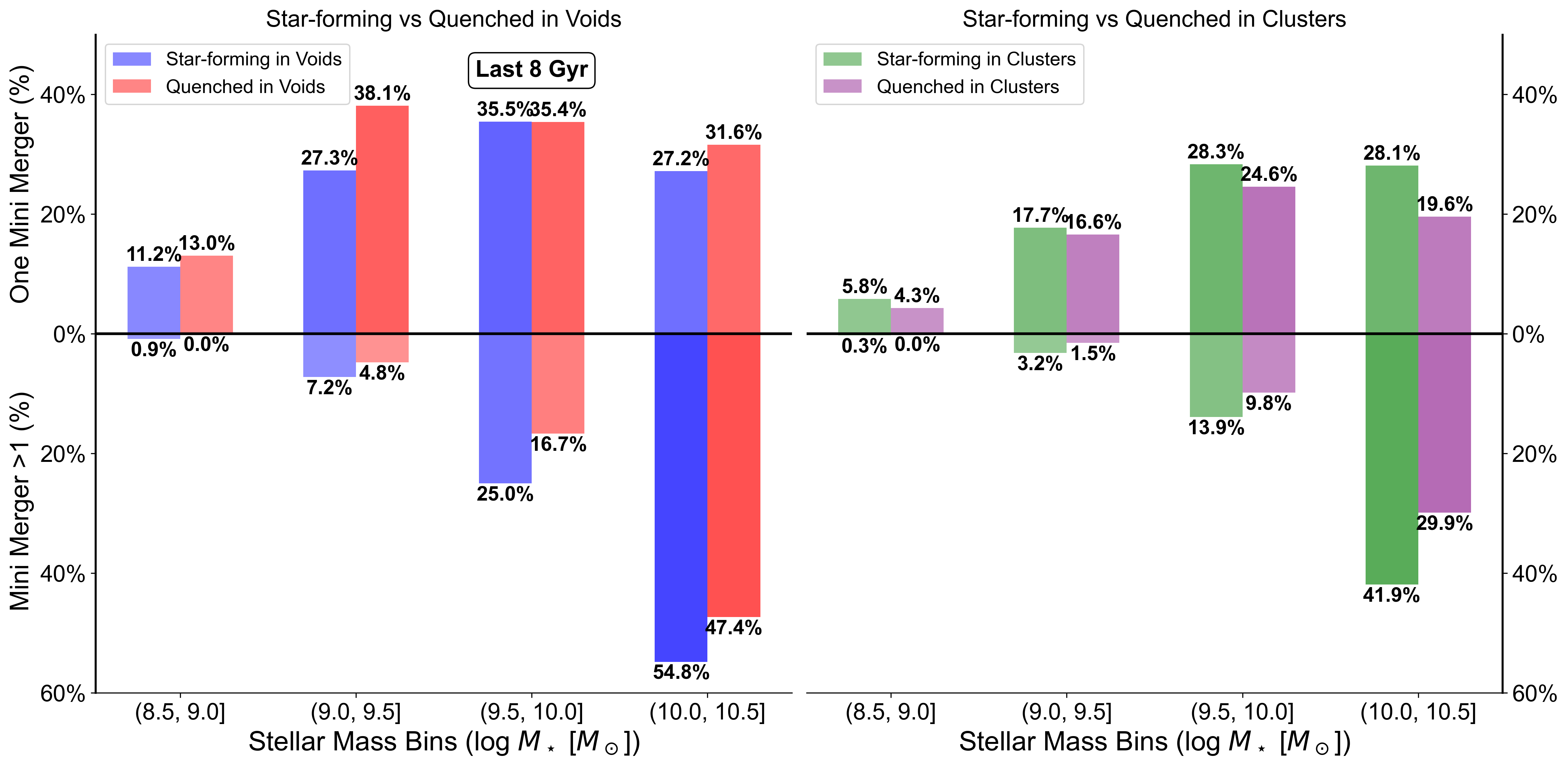}
    
    \caption{
    Merger fractions (major, minor, and mini) over the last 8~Gyr for star-forming and quenched galaxies in clusters and voids. 
    In each bar plot, the upper bars represent the percentage of galaxies within each stellar-mass bin that have experienced exactly one merger, 
    while the lower bars indicate the percentage of galaxies that have undergone multiple mergers 
    ($N_{\mathrm{mergers}} > 1$).
    }
    \label{fig:01stellar_mass_comparison011}
\end{figure*}

\section{Merger fractions across stellar-mass bins}
\label{appendix:Merger fractions across stellar-mass bins}

To provide a more detailed view of merger activity across different stellar-mass ranges, we present in Figure~\ref{fig:01stellar_mass_comparison011} the full comparison of major, minor, and mini merger fractions for star-forming and quenched galaxies in void and cluster environments over the last 8~Gyr. Only the key trends are summarized in the main text; the complete breakdown is included here for completeness. By separating galaxies into stellar-mass bins and environmental categories, we provide a more detailed view of how merger activity evolves with cosmic time. 

Overall, the extended figures support the robustness of the conclusions drawn in the main analysis.Major mergers are consistently more common in star-forming galaxies than in quenched systems across most stellar-mass bins, with slightly higher fractions in void environments. Minor mergers show a comparable but weaker trend, with fractions that generally increase toward higher stellar masses, particularly for cluster galaxies. In contrast, mini mergers exhibit the strongest dependence on stellar mass, becoming increasingly dominant at the high-mass end in both environments. Overall, these distributions indicate that while major and minor merger fractions vary moderately with environment and galaxy type, mini mergers represent the most frequent merger channel, especially for massive galaxies.

\section{Last mergers: star-forming versus quenched galaxies within each environment}
\label{appendix:last_merger_sf_q}

\begin{figure*}[t]
    \centering
    \includegraphics[width=0.9\textwidth, keepaspectratio]{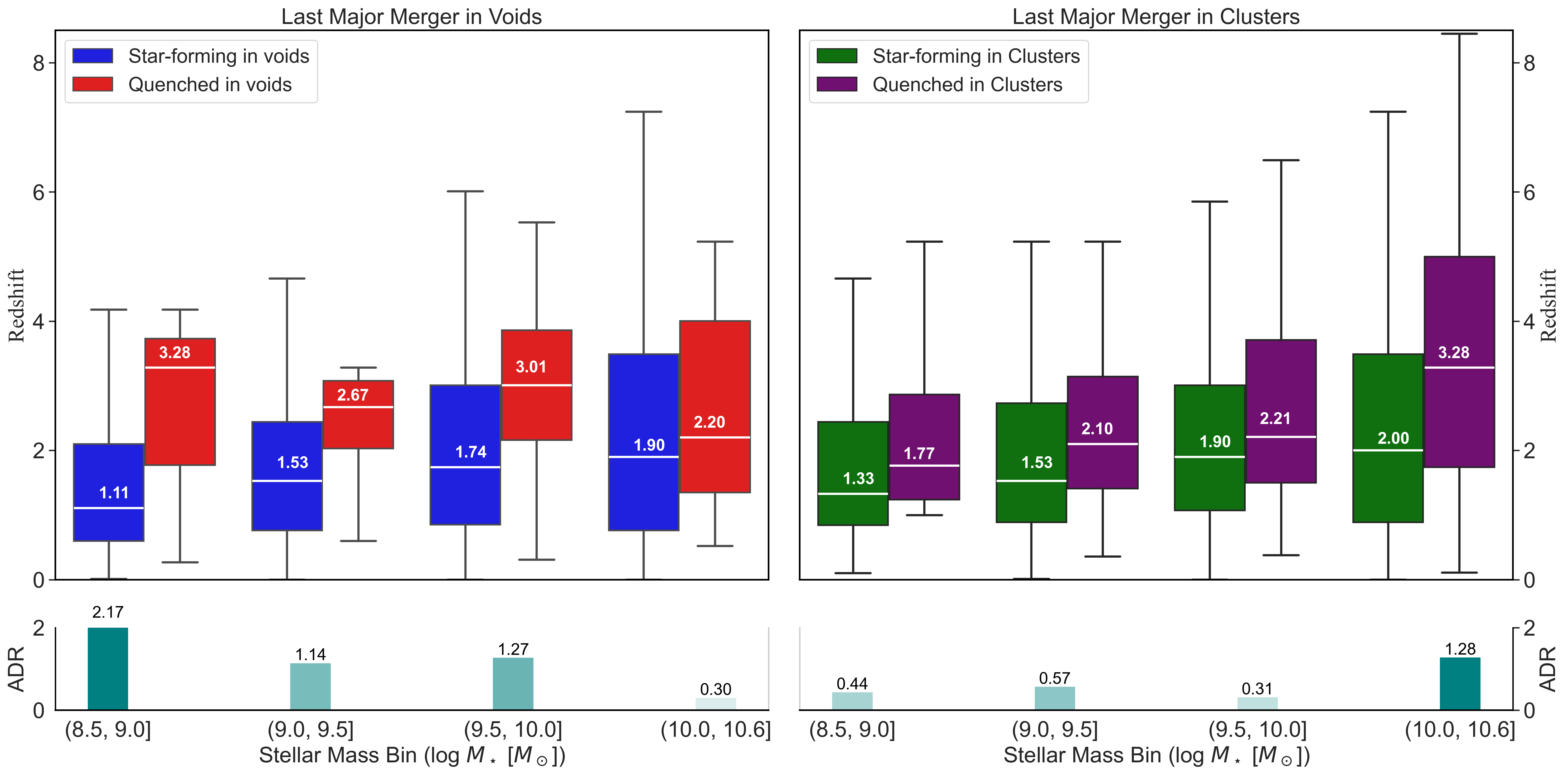}

    \vspace{0.5em}

    \includegraphics[width=0.9\textwidth, keepaspectratio]{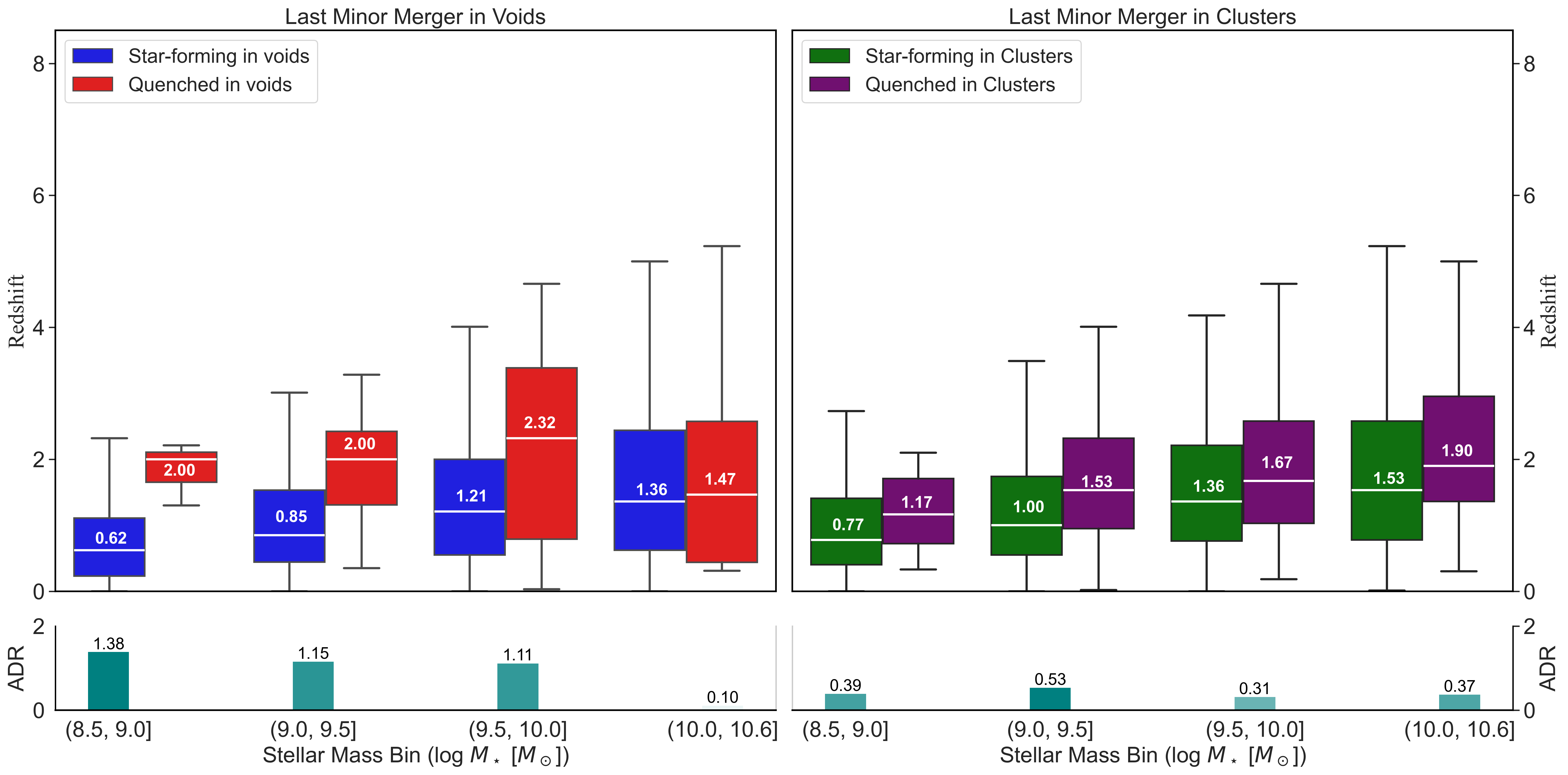}
    \caption{Comparison of last major (top) and minor (bottom) mergers between star-forming and quenched galaxies within voids (left) and within clusters (right), illustrated through box plots of redshift distributions and bar charts displaying absolute differences in redshift (ADR) between star-forming and quenched galaxies across stellar mass bins in each environment. Each horizontal white line in the box plot represents the median value of the last merger in redshift for each stellar mass bin. This figure complements Fig.~\ref{fig:01stellar_mass_comparison010} in the main text, which instead compares void and cluster environments directly within each galaxy type.}
    \label{fig:last_merger_appendix}
\end{figure*}

For completeness, Figure~\ref{fig:last_merger_appendix} shows the complementary comparison between star-forming and quenched galaxies within each environment (voids and clusters shown separately), as opposed to the void-cluster comparison for each galaxy type presented in the main text (Fig.~\ref{fig:01stellar_mass_comparison010}). The bar charts display the absolute differences in redshift (ADR) between star-forming and quenched galaxies within each environment. Overall, star-forming galaxies, regardless of environment, underwent their last mergers more recently than quenched galaxies, consistent with the earlier stellar mass assembly of quenched systems discussed in Sect.~3.3. The ADR between star-forming and quenched galaxies is more significant in voids than in clusters for both major and minor mergers, indicating a more pronounced temporal gap between the merger histories of these two populations in low-density environments.

\end{document}